\documentclass[a4paper,11pt]{article}
\pdfoutput=1 

\usepackage{jcappub} 

\usepackage[T1]{fontenc} 
\usepackage{enumitem}
\usepackage{verbatim}
\usepackage[T1]{fontenc}
\usepackage[utf8]{inputenc}
\usepackage[american]{babel}
\usepackage{epsfig}
\usepackage{graphicx,subcaption,caption}
\usepackage{booktabs}
\usepackage{multirow}
\usepackage{dcolumn}
\usepackage{amsmath}
\usepackage{mathtools}
\usepackage{amsfonts}
\usepackage{amssymb}
\usepackage{ulem}
\usepackage{epstopdf}
\usepackage{bm}
\usepackage{siunitx}
\usepackage{braket}
\usepackage{enumitem}
\usepackage{soul}

\title{Inflationary Potential as seen from Different Angles: Model Compatibility from Multiple CMB Missions}

\author[a,1]{William Giar\`e, \note{Corresponding author.}}
\author[b,c]{Supriya Pan,}
\author[a]{Eleonora Di Valentino,}
\author[d]{Weiqiang Yang,}
\author[e]{Jaume de Haro,}
\author[f]{Alessandro Melchiorri}

\affiliation[a]{School of Mathematics and Statistics, University of Sheffield, Hounsfield Road, Sheffield S3 7RH, United Kingdom}
\affiliation[b]{Department of Mathematics, Presidency University, 86/1 College Street, Kolkata 700073, India}
\affiliation[c]{Institute of Systems Science, Durban University of Technology, PO Box 1334, Durban 4000, Republic of South Africa}
\affiliation[d]{Department of Physics, Liaoning Normal University, Dalian, 116029, P. R. China}
\affiliation[e]{Departament of Mathematics, Polytechnic University of Catalonia, Diagonal 647, 08028 Barcelona, Spain}
\affiliation[f]{Physics Department and INFN, Universit\`a di Roma ``La Sapienza'', Ple Aldo Moro 2, 00185, Rome, Italy}

\emailAdd{w.giare@sheffield.ac.uk}
\emailAdd{supriya.maths@presiuniv.ac.in}
\emailAdd{e.divalentino@sheffield.ac.uk}
\emailAdd{d11102004@163.com}
\emailAdd{jaime.haro@upc.edu}
\emailAdd{alessandro.melchiorri@roma1.infn.it}

\abstract{The cosmic microwave background (CMB) temperature and polarization anisotropies, as observed by independent astronomical missions such as WMAP, Planck, and most recently the Atacama Cosmology Telescope and the South Pole Telescope have played a vital role in accurately constraining cosmological theories and models, establishing cosmic inflation as the most widely accepted theory for describing the physics of the early Universe. However, the absence of a definitive detection of B-mode polarization and the emerging discrepancies among different CMB experiments present a challenge in determining which inflationary models best explain the observed data. In this work, we further explore this difficulty and conduct a case study by analyzing four well-known inflationary potentials in light of the latest CMB temperature and polarization anisotropy measurements and lensing data released by the Planck satellite and the Atacama Cosmology Telescope. Additionally, we incorporate B-modes polarization data from the BICEP/Keck Collaboration, as well as Baryon Acoustic Oscillations and Redshift Space Distortions measurements from BOSS DR12 and eBOSS DR16. We show that the most typical models such as Starobinsky and $\alpha$-attractors are in disagreement with the Atacama Cosmology Telescope small-scale CMB measurements, particularly when combined with B-modes polarization data. On the other hand, these potentials are in perfect agreement with the Planck measurements at larger angular scales. This dichotomy makes it challenging to identify a single model or a group of models that can be universally considered as the preferred choice based on all available CMB observations.}

\begin{document}
\maketitle
\flushbottom

\section{Introduction}

The theory of cosmic inflation~\cite{Guth:1980zm,Sato:1980yn,Linde:1981mu} was proposed as a heuristic solution to the shortcomings of big bang cosmology, such as the flatness problem, horizon problem, monopole problem, and the origin of large-scale structure in the Universe.\footnote{It is worth noting that a specific inflationary model known as the $R^2$ inflation (where $R$ represents the Ricci scalar) was proposed by Starobinsky in 1980~\cite{Starobinsky:1980te}, a year earlier of~\cite{{Guth:1980zm,Sato:1980yn}}. However, unlike Guth's paper~\cite{Guth:1980zm}, the salient features of inflation and its motivation were not directly pointed out in~\cite{Starobinsky:1980te}. It is anyway worth nothing that effects of a Higgs field and the exponential expansion were also discussed earlier in \cite{1980ApJ...239..428K}.}  Over the past few decades, inflation has garnered significant support thanks to its ability to elegantly explain the tiny anisotropies observed in the cosmic microwave background (CMB) radiation, which have been measured with increasingly precision by various astronomical missions such as  the COBE satellite~\citep{Fixsen:1993rd, Bennett:1996ce}, the WMAP satellite~\citep{Hinshaw:2012aka,WMAP:2012fli,WMAP:2003syu} and, more recently, the Planck satellite~\citep{Planck:2019nip,Planck:2018nkj,Planck:2013jfk,Planck:2015sxf,Planck:2018jri}, the Atacama Cosmology Telescope~\citep{ACT:2020frw,ACT:2020gnv,ACT:2023kun} and the South Pole Telescope~\citep{SPT-3G:2014dbx,SPT-3G:2021eoc,SPT-3G:2022hvq}. Although the inflationary paradigm remains strong and in very good health, it is important to emphasize that a wide range of inflationary models can emerge from very different fundamental mechanisms or be motivated by various phenomenological considerations ranging from modified gravity theories, high-energy particle physics, and supersymmetric frameworks\footnote{See, \textit{e.g.}, Refs.~\cite{Martin:2013tda,Lyth:1998xn,Linde:2007fr,Baumann:2014nda} and references therein for reviews and discussions.}. As a result, the nature of the inflation field (or fields) still remains elusive, leaving the integration of inflation into a more fundamental theory an active field of research with interesting ongoing debates.

In this regard, over the last several years, 
numerous inflationary models and theories have been tested against a wide range of available data, including CMB, Big Bang Nucleosynthesis (BBN), and Gravitational Wave measurement, see \textit{e.g.}~\cite{Leach:2002ar,Boubekeur:2005zm,Martin:2006rs,Moss:2007qd,Bezrukov:2010jz,Zhao:2011zb,Martin:2013nzq,Martin:2014rqa,Martin:2014lra,Carrillo-Gonzalez:2014tia,Creminelli:2014oaa,DiValentino:2016nni,DiValentino:2016ziq,Campista:2017ovq,Giare:2019snj,Forconi:2021que,Dai:2019ejv,Baumann:2015xxa,Odintsov:2020ilr,Giare:2020plo,Oikonomou:2021kql,Odintsov:2022cbm,Namba:2015gja,Peloso:2016gqs,Pi:2019ihn,Ozsoy:2020ccy,Stewart:2007fu, Mukohyama:2014gba,Giovannini:2015kfa,Giovannini:2018dob,Giovannini:2018nkt,Giovannini:2018zbf,Giare:2020vhn,Giare:2020vss,Giare:2020plo,Giare:2022wxq,Baumgart:2021ptt,Franciolini:2018ebs,DEramo:2019tit,Giare:2019snj,Caldwell:2018giq,Clarke:2020bil,Caprini_2018,Giare:2020vhn,Allen:1997ad,Smith:2006nka,Boyle:2007zx,Kuroyanagi:2014nba,Ben-Dayan:2019gll,Aich:2019obd,Cabass:2015jwe,Vagnozzi:2020gtf,Benetti:2021uea,Calcagni:2020tvw,Oikonomou:2022ijs,Barrow:1993ad,Peng:2021zon,Ota:2022hvh,Odintsov:2022sdk,Baumgart:2021ptt,Capurri:2020qgz,Canas-Herrera:2021sjs,Odintsov:2023aaw,Oikonomou:2023bah,Fronimos:2023tim,Fronimos:2023tim,Cai:2022lec,Oikonomou:2022irx,Gangopadhyay:2022vgh,Odintsov:2022hxu,Odintsov:2022cbm,Odintsov:2020mkz,Galloni:2022mok,Braglia:2022phb,Giare:2023kiv,Antoniadis:2023zhi,NANOGrav:2023gor,NANOGrav:2023hvm,Vagnozzi:2023lwo,Oikonomou:2023qfz}  and the references therein. As the observational data are becoming increasingly sensitive, the constraints on the inflationary models have improved, leading to reduced uncertainties. This progress is evident in the constraints on inflationary parameters derived from the CMB by Planck 2018 final release~\cite{Planck:2018jri}
compared to the constraints obtained from Planck 2015~\cite{Planck:2015sxf}. Moreover, we have now access to CMB temperature and polarization anisotropies data from different independent CMB experiments such as the Atacama Cosmology Telescope~\cite{ACT:2020frw,ACT:2023kun} and the South Pole Telescope~\cite{SPT-3G:2022hvq,SPT-3G:2021eoc}. This availability of multipole CMB measurements presents an exciting opportunity to comprehensively test the results obtained for different inflationary scenarios by multiple independent CMB experiments, enabling us to better understand which inflationary models can be favored or disfavored in light of a wide range of observations probing different angular scales.

However, and despite the best efforts, determining the definitive inflationary model still remains an open question~\cite{Martin:2013nzq}. The absence of detection of B-mode polarization in the cosmic microwave background~\cite{BICEP:2021xfz}, which would represent a smoking-gun evidence for inflation, offering valuable information on its underlying micro-physics, presents a major challenge. In addition, the emerging disagreements among independent CMB observations~\cite{Lin:2019zdn,Handley:2020hdp,LaPosta:2022llv,DiValentino:2022rdg,DiValentino:2022oon,Giare:2022rvg,Calderon:2023obf,Giare:2023xoc} often involve inflationary parameters~\cite{Forconi:2021que,Giare:2022rvg,DiValentino:2022oon}, contributing to the ongoing uncertainty. Thus, with inflation, one of the important cornerstones of modern cosmology, we find ourselves between the known and unknown. 

In this article we delve deeper into this difficulty. As a case study, we consider 
four inflationary models and constrain them in light of the latest CMB anisotropies and lensing data released by two independent experiments: the Planck satellite and the Atacama Cosmology Telescopen (ACTPol). We consider these experiments in conjunction with B-modes polarization data from the BICEP/Keck Collaboration, and Baryon Acoustic Oscillations (BAO) and Redshift Space Distortions (RSD) measurements from BOSS DR12 and eBOSS DR16. Our primary goal is to gain insight into how the emerging disagreements between distinct CMB observations can be reflected in the constraints on inflationary models and study their implications for model selection. Additionally, we actively pursue an answer to the question of whether there exists an inflationary model (or a class of models) that broadly aligns with all available CMB measurements.
The article is structured as follows. In \autoref{sec-inflation-models} we introduce the inflationary models that we wish to study in this work. \autoref{sec-methods-data} describes the statistical methods and the observational data used to constrain the models. Then in \autoref{sec-results}, we discuss the main results extracted out of the present inflationary models. Finally, we close the article by presenting a brief summary of the entire work in \autoref{sec-conclu}. 

\section{Inflationary Models}
\label{sec-inflation-models}

In this Section we  present some inflationary models that we have confronted with the observational data. For each
model, we provide a concise overview of physical properties and derive relationships between observables. These relationships will be subsequently assumed in the data analysis, as explained in the following section. As for the mathematical set-up, we consider a homogeneous and isotropic Universe described by the spatially flat Friedmann-Lema\^{i}tre-Robertson-Walker (FLRW) line element  $ds^2 = dt^2 + a^2 (t) (dx^2 + dy^2 +dz^2)$ where $(t, x, y, z)$ are the co-moving coordinates and $a(t)$ is the expansion scale factor of the Universe.   

\subsection{Starobinsky inflation}

One of the most successful inflationary proposals is the Starobinsky model~\cite{Starobinsky:1980te}, which comes from the modified $f(R)$-gravity, being $R$ the Ricci scalar. Choosing  $f(R) = R + R^2/m^2$, where the parameter $m$ has the unit of mass, one can see that the model contains an inflationary period at early times and ends in an oscillating phase producing particles named by Starobinsky {\it scalarons} which thermalize the Universe  in a matter domination era~\cite{Amoros:2014tha}. Effectively, for this model, the modified Friedmann equation is given by
\begin{eqnarray}
    H^2=-\frac{12}{m^2}\left(3\dot{H}H^2+H\ddot{H}-\frac{1}{2}\dot{H}^2 \right),
\end{eqnarray}
and during the slow-roll phase, $\dot{H}\ll H^2$, will become
\begin{eqnarray}
    \ddot{H}=-3H\dot{H}-\frac{m^2H}{12},
\end{eqnarray}
which, for $t<t_{end}$ have the following solution
\begin{eqnarray}
    H(t)=\frac{m^2(t_{end}-t)}{36}\Longrightarrow a(t)=a_{end}e^{-\frac{18H^2(t)}{m^2}},
\end{eqnarray}
where the subindex $"end"$ denotes the end of the inflation. Thus, denoting by $"*"$ the scale of inflation, in order  to solve the Big Bang shortcomings such as the horizon and flatness problems, one needs around $60$ $e$-folds, which imposes  $H_*\sim m$, see also \autoref{appendix} for a more detailed discussion.

On the other hand, at late times one can show that the scale factor evolve as~\cite{Starobinsky:1980te} (see also 
\cite{Amoros:2014tha} for an explicit deduction)
\begin{eqnarray}
    a(t)=t^{2/3}\left( 1+\frac{2}{3}\frac{\sin(tm/\sqrt{6}  )}{tm/\sqrt{6}}  \right)\cong t^{2/3}.
    \end{eqnarray}

So far we are working in the Jordan frame, and to pass to the Einstein one, we have  to use the following transformation
\begin{eqnarray}
    \varphi=\sqrt{\frac{3}{2}}\ln(f'(R))M_{pl},\quad V(\varphi)=\frac{Rf'(R)-f(R)}{2(f'(R))^2}M_{pl}^2,
\end{eqnarray}
where $M_{pl}$ denotes the reduced Planck's mass.
In the particular case of  the Starobinsky $f(R)$-gravity  model, the potential becomes 
\begin{eqnarray}
    V (\varphi) = \frac{M_{pl}^2 m^2}{8} \left(1 - e^{-\sqrt{\frac{2}{3}}\varphi/M_{pl}}\right)^2.
    \label{eq:pot_Starob}
\end{eqnarray}

To achieve a relation among the number of last $e$-folds $\mathcal{N}$ (from the horizon crossing to the end of inflation),  the spectral index of the scalar perturbations ($n_s$), its running ($\alpha_s$) and the tensor-to-scalar ratio ($r$), we define the slow roll parameters:
\begin{eqnarray}
    \epsilon=\frac{M_{pl}^2}{2}\left(\frac{V'}{V} \right)^2, \quad \eta={M_{pl}^2}\frac{V''}{V} \quad \mbox{and}\quad
    \xi=M_{pl}^4\left( \frac{V'V'''}{V^2}\right),
    \end{eqnarray}
Regardless of the specific model assumed, these
measureable parameters can be written in terms of the slow roll parameters at the horizon crossing as: 
\begin{eqnarray}
    r=16\epsilon_*, \quad 1-n_s=6\epsilon_*-2\eta_*\quad \mbox{and} \quad \alpha_s=-2\xi_*+16\epsilon_*\eta_*-24\epsilon_*^2,
\end{eqnarray}
Assuming the Starobinsky potential, Eq.~\eqref{eq:pot_Starob}, at the horizon crossing we get
\begin{eqnarray}
    \epsilon_*\cong \frac{4}{3}e^{-2\sqrt{\frac{2}{3}}\varphi_*/M_{pl}}, \quad \eta_*\cong -\frac{4}{3}e^{-\sqrt{\frac{2}{3}}\varphi_*/M_{pl}}
    \quad\mbox{and}\quad \xi_*\cong  \frac{16}{9}e^{-2\sqrt{\frac{2}{3}}\varphi_*/M_{pl}},   \end{eqnarray}
and we can see that, for this model,  $1-n_s\cong -2\eta_*$ and $\alpha_s\cong -2\xi_*$. On the other hand, the number of last $e$-folds 
can be computed as 
\begin{eqnarray}\label{N_0}
{\mathcal N}=\frac{1}{M_{pl}}\int_{\varphi_*}^{\varphi_{end}}\frac{1}{\sqrt{2\epsilon}}d\varphi
\cong\frac{3}{4}e^{\sqrt{\frac{2}{3}}\varphi_*/M_{pl}}.
\end{eqnarray}
By combining the previous relations, we can immediately obtain a set of equations linking together the inflationary parameter and the last $e$-folds of expansion $\mathcal N$:
\begin{eqnarray}
  1-n_s\cong \frac{2} {\mathcal N},  \qquad \alpha_s\cong -\frac{2}{{\mathcal N}^2}, \qquad r\cong \frac{12}{{\mathcal N}^2}.
    \label{eq.Starob}
\end{eqnarray}

\subsection{$\alpha$-Attractors}

The $\alpha$-Attractors were introduced for the first time in Ref.~\cite{Kallosh:2013yoa}, obtaining  models which generalize  the  Starobinsky model (see also~\cite{Kehagias:2013mya} for a detailed revision of the Starobinsky model).

Here, we consider an $\alpha$-Attractor in the context of Quintessential Inflation~\cite{Peebles:1998qn}. For this reason we deal with 
the following Lagrangian motivated by supergravity and corresponding to a non-trivial K\"ahler manifold (see for instance~\cite{Dimopoulos:2017zvq} and the  references therein), combined with  a standard   exponential potential,
\begin{eqnarray}\label{lagrangian}
\mathcal{L}=\frac{1}{2}\frac{\dot{\phi}^2}{(1-\frac{\phi^2}{6\alpha}  )^2}M_{pl}^2-\lambda M_{pl}^4 e^{-\kappa \phi},
\end{eqnarray}
where $\phi$ is a dimensionless scalar field; $\alpha$, $\kappa$ and $\lambda$ are positive dimensionless constants.

In order that the kinetic term has the canonical  form, one can redefine the scalar field as follows,
\begin{eqnarray}
\phi= \sqrt{6\alpha}\tanh\left(\frac{\varphi}{\sqrt{6\alpha}M_{pl}}  \right),
\end{eqnarray}
obtaining the following potential~\cite{AresteSalo:2021wgb},
\begin{eqnarray}\label{alpha}
V(\varphi)=\lambda M_{pl}^4e^{-n\tanh\left(\frac{\varphi}{\sqrt{6\alpha}M_{pl}} \right)},
\end{eqnarray}
where we have introduced the dimensionless parameter $n = \kappa\sqrt{6\alpha}$. 

The asymptotic values of the potential are $V_{\pm} = \lambda M_{pl}^4\exp(\pm n)$, and the  
slow-roll parameters  evaluated at the horizon crossing, which will happen for large values of 
$\cosh\left(\frac{\varphi}{\sqrt{6\alpha}M_{pl}} \right)$, are 
\begin{eqnarray}\label{parameters}
\epsilon_*=\frac{n^2}{12\alpha}\frac{1}{\cosh^4\left( \frac{\varphi_*/M_{pl}}{\sqrt{6\alpha}}\right)}, \quad \eta_*
\cong -\frac{n}{3\alpha}\frac{1}{\cosh^2\left(\frac{\varphi_*/M_{pl}}{\sqrt{6\alpha}} \right)},
\quad {\xi_*\cong \frac{n^2}{9\alpha^2}\frac{1}{\cosh^4\left(\frac{\varphi_*/M_{pl}}{\sqrt{6\alpha}} \right)}}, 
\end{eqnarray}
with $\varphi_*<0$.  Next, we calculate the number of last $e$-folds from the horizon crossing to the end of inflation, which for small values of $\alpha$ is given by
\begin{eqnarray}\label{N}
{\mathcal N}=\frac{1}{M_{pl}}\int_{\varphi_*}^{\varphi_{end}}\frac{1}{\sqrt{2\epsilon}}d\varphi
\cong \sqrt{\frac{3\alpha}{4\epsilon_*}},
\end{eqnarray}
so we get the standard form of the spectral index, its running and the tensor/scalar ratio for an $\alpha$-Attractor as follows
\begin{eqnarray}\label{power}
1-n_s\cong 
\frac{2}{{\mathcal N}}, \qquad 
{\alpha_s\cong -\frac{2}{{\mathcal N}^2}},
\qquad
r\cong 16\epsilon_*\cong\frac{12\alpha}{{\mathcal N}^2}.
\qquad 
\label{eq.alphaAtt}
\end{eqnarray}

\subsection{Polynomial inflation}

We start postulating  the following dynamics~\cite{deHaro:2016cdm}: 
\begin{eqnarray}
    \dot{H}=\left\{\begin{array}{ccc}
    (-3H_{kin}^2+\Lambda)\left( \frac{H}{H_{kin}}\right)^{\alpha} & \mbox{for}   & H\geq H_{kin}  \\
        -3H^2+\Lambda & \mbox{for} & H\leq H_{kin}, \end{array}
\right.
\end{eqnarray}
where $H_{kin}$ is the value of the Hubble rate at the beginning of kination (a regime where all the energy is kinetic), $\Lambda\ll H_{kin}$ is a cosmological constant and $\alpha\in [0,1]$ is the parameter which defines the family of models under consideration.
Note that the value of $H_{kin}$ can be calculated from the value of the 
slow-roll parameter $\epsilon$ at the horizon crossing, whose value is
\begin{eqnarray}
    \epsilon_*=3\left(\frac{H_{kin}}{H_*} \right)^{2-\alpha}=\frac{1-n_s}{4-\alpha},
\end{eqnarray}
and the power spectrum of scalar perturbations (\ref{spectrum}), obtaining
\begin{eqnarray}
    H_{kin}\sim 7\times 10^{-4}\left(\frac{1-n_s}{3(4-\alpha)}  \right)^{\frac{4-\alpha}{2(2-\alpha)}}M_{pl},
\end{eqnarray}
which for the central value of the spectral index and $0\leq \alpha\leq 1$, leads to 
\begin{eqnarray}
    10^{-7} M_{pl} \lesssim H_{kin}\lesssim 2\times 10^{-6} M_{pl}.
\end{eqnarray}

On the other hand,
the corresponding effective Equation of State (EoS) parameter $w_{\rm{eff}}=-1-\frac{2\dot{H}}{3H^2}$ is given by
\begin{eqnarray}
    w_{\rm{eff}}=\left\{\begin{array}{ccc}
   -1+2 \left(1-\frac{\Lambda}{3H_{kin}^2} \right)\left( \frac{H}{H_{kin}}\right)^{\alpha-2} & \mbox{for}   & H\geq H_{kin}  \\
\left(1-\frac{2\Lambda}{3H_{kin}^2} \right)& \mbox{for} & H\leq H_{kin}, \end{array}
\right.
\end{eqnarray}
which shows that for $H\gg H_{kin}$ one has $w_{eff}\cong -1$ (early slow-roll phase). When $H\cong H_{kin}$, the EoS
parameter satisfies $w_{\rm{eff}}\cong 1$ (kination stage), and finally, for 
$H\cong \sqrt{\frac{\Lambda}{3}}$
one also has $w_{\rm{eff}}\cong -1$ (late quasi de Sitter period).

The corresponding potential can be reconstructed from the Raychaudhuri equation 
\begin{eqnarray}
    \varphi=M_{pl}\int \sqrt{-2\dot{H}}dt=-M_{pl}\int\sqrt{\frac{-2}{\dot{H}}}dH,
\end{eqnarray}
what leads the relation between the inflaton field and the Hubble rate, and once one has this relationship, we can use the formula 
\begin{eqnarray}
    V(\varphi)=\left(3H^2(\varphi)+\dot{H}(\varphi)   \right)M_{pl}^2,
\end{eqnarray}
to obtain

\begin{eqnarray}\label{QI-1}
 V(\varphi)=\left\{\begin{array}{cc}
           3\left(\frac{H_{kin} M_{pl}}{\varphi_{kin}} \right)^2\left( \frac{\varphi}{\varphi_{kin}}\right)^{\frac{2\alpha}{2-\alpha}}
           \left[\varphi^2-\varphi_{kin}^2\left(1- \frac{\Lambda}{3H_{kin}^2} \right)\right]& \varphi\leq \varphi_{kin}\\ \\
          \Lambda M_{pl}^2 & \varphi\geq \varphi_{kin},
          \end{array} \right.
\end{eqnarray}

where $\varphi_{kin}\equiv -\frac{2\sqrt{2}}{\sqrt{3}(2-\alpha)}
\frac{H_{kin}M_{pl}}{\sqrt{H_{kin}^2-\frac{\Lambda}{3}}}\cong
-\frac{2\sqrt{2}}{\sqrt{3}(2-\alpha)}M_{pl}$.

Notice that for $\alpha=0$, the potential is quadratic; $\alpha=\frac{2}{3}$ a cubic potential is recovered and for $\alpha=1$, it represents a quartic potential.  Lastly, note  that for a polynomial potential of the form  $V(\varphi)=\lambda \left(\frac{\varphi}{M_{pl}} \right)^{2n}$, one has 
\begin{eqnarray}
    \epsilon_*=2n^2\frac{M_{pl}^2}{\varphi^2_*}, \quad 
     \eta_*=2n(2n-1)\frac{M_{pl}^2}{\varphi^2_*}, \quad  
     {\xi_*=8n^2(n-1)(2m-1)\frac{M_{pl}^4}{\varphi^4_*}, }
     \end{eqnarray}
and thus, 
\begin{eqnarray}
    1-n_s= 4n(n+1)\frac{M_{pl}^2}{\varphi^2_*}, \quad 
    r= 32n^2 \frac{M_{pl}^2}{\varphi^2_*}, 
    \quad { \alpha_s=-16n^2(n+1)\frac{M_{pl}^4}{\varphi^4_*} }.    
    \end{eqnarray}

On the other hand, taking into account that $\varphi_{end}=\sqrt{2}nM_{pl}$,  the number of last $e$-folds is
\begin{eqnarray}
    {\mathcal N}=\frac{\varphi^2_*}{4nM_{pl}^2}-\frac{n}{2},
    \end{eqnarray}
and consequently, we find 
\begin{eqnarray}
1-n_s=\frac{2(n+1)}{2{\mathcal N}+n},\quad
 \alpha_s=-
 \frac{4(n+1)}{(2{\mathcal N}+n)^2}, \quad
  r=\frac{16n}{2{\mathcal N}+n}.
 \label{eq.PolInf}
\end{eqnarray}

\subsection{SUSY inflation}

Here we consider an Exponential SUSY Inflation-type potential, coming from a 
K\"{a}lher potential (see \cite{Martin:2013tda} and the references therein) given by
\begin{eqnarray}\label{QI-2-SUSY}
V(\varphi)= \lambda M_{pl}^4 \left( 1-e^{\varphi/M_{pl}} \right), 
\end{eqnarray}
where the parameter $\lambda\sim 2\times 10^{-9}$  is obtained using the power spectrum of scalar perturbations in \autoref{spectrum}.  For this model one has
\begin{eqnarray}
    \epsilon_*\cong \frac{1}{2}e^{2\varphi_*/M_{pl}}, \quad \eta_*\cong 
-e^{\varphi_*/M_{pl}},  \quad { \xi_*\cong e^{2\varphi_*/M_{pl}}},
    \end{eqnarray}
and thus 
\begin{eqnarray}
    1-n_s\cong -2\eta_*\cong -2e^{\varphi_*/M_{pl}}, \qquad { \alpha_s\cong -2\xi_*\cong -2e^{2\varphi_*/M_{pl}}}.
\end{eqnarray}

On the other hand, the number of last  $e$-folds is given by
\begin{eqnarray}
    {\mathcal N}\cong e^{-\varphi_*/M_{pl}}.
\end{eqnarray}
which leads to the following relations
\begin{eqnarray}
 1-n_s \cong \frac{2}{\mathcal N}, \qquad
  \alpha_s\cong -\frac{2}{{\mathcal N}^2}, \qquad r\cong \frac{8}{{\mathcal N}^2}. 
  \label{eq.SUSY}
\end{eqnarray}

Therefore,  the Starobinsky, $\alpha$-Attractor and exponential SUSY models lead to  the same relationship between the number of last $e$-folds,  the spectral index and its running while differing for the amplitude of primordial tensor modes.

\section{Methods and data}
\label{sec-methods-data}

A typical approach used when constraining inflationary parameters with cosmological and astrophysical observations is to remain agnostic about the specific shape of the inflationary potential and adopt a generic power-law form for the power spectrum of primordial adiabatic components, given by:
\begin{equation}
\log \mathcal{P}_{\rm s}(k) = \log A_{\mathrm{s}} + \left(n_{\mathrm{s}}-1\right) \log \left(\frac{k}{k_{\star}}\right) + \frac{1}{2} \alpha_s \log^2 \left(\frac{k}{k_{\star}}\right)
\label{PLS}
\end{equation}
where $k_{\star}$ is the pivot scale (in this work fixed to $k_{\star}=0.05$ Mpc$^{-1}$), while the amplitude of the spectrum $A_{\mathrm{s}}$, the spectral index $n_s$, and eventually the running $\alpha_s$ are regarded as independent free parameters to be constrained by data. Similarly, primordial gravitational waves are usually characterized by a power-law spectrum
\begin{equation}
\log \mathcal{P}_{\rm T}(k) = \log (A_{\mathrm{T}}) + n_{\rm T} \log \left(\frac{k}{k_{\star}}\right) + \dots 
\label{PLT}
\end{equation}
whose amplitude is parametrized in terms of the so-called tensor-to-scalar ratio $r\equiv \mathcal P_{\rm T} (k_{\star})/ \mathcal P_{\rm S}(k_{\star})$ and whose tilt $n_{\rm T}=-r/8$ is fixed by the slow-roll consistency relation, leaving only one free degree of freedom. Once the free parameters are constrained by data, the model comparison is subsequently performed by studying how well the predictions of generic inflationary potentials align with such model-independent constraints.

The former approach is clearly motivated by the fact that in single-field slow-roll inflation, the spectra of scalar and tensor perturbations can be well approximated by the above mentioned power-law forms, regardless of the specific model assumed. However, when assuming a specific inflationary potential, the parameters appearing in these power-laws are no longer independent of each other and additional relations linking together the spectral index $n_{s}$, its running $\alpha_{s}$, the tensor amplitude $r$ and  the number of last $e$-folds  ($\mathcal{N}$) can be derived, thereby reducing the number of free degrees of freedom and facilitating a more accurate analysis. Hence, an alternative approach is to start with a specific inflationary model and impose the theoretical relations within the cosmological framework to conduct a more focused investigation of its properties.

\begin{table}[t!]
\begin{center}
\renewcommand{\arraystretch}{1.3}
\resizebox{\textwidth}{!}{
\begin{tabular}{l c c c c c c c c c c c c c c c }
\hline
\textbf{Parameter} & \textbf{ $\Lambda$CDM } & \textbf{ Starobinsky } & \textbf{$\alpha$ - Attractor } & \textbf{Polynomial inflation} & \textbf{SUSY} \\ 
\hline\hline

$ \Omega_\mathrm{b} h^2  $ & $[0.005\,,\,0.1]$ & $[0.005\,,\,0.1]$ & $[0.005\,,\,0.1]$ & $[0.005\,,\,0.1]$ & $[0.005\,,\,0.1]$ \\ 
$ \Omega_\mathrm{c} h^2  $ & $[0.001\,,\,0.99]$ & $[0.001\,,\,0.99]$ & $[0.001\,,\,0.99]$ & $[0.001\,,\,0.99]$ & $[0.001\,,\,0.99]$ \\ 
$ 100\theta_\mathrm{MC}  $ & $[0.5\,,\,10]$ & $[0.5\,,\,10]$ & $[0.5\,,\,10]$ & $[0.5\,,\,10]$ & $[0.5\,,\,10]$ \\ 
$ \tau  $ & $[0\,,\,0.8]$ & $[0\,,\,0.8]$ & $[0\,,\,0.8]$ & $[0\,,\,0.8]$ & $[0\,,\,0.8]$ \\
$ \log(10^{10} A_\mathrm{s})  $ & $[1.6\,,\,3.9]$ & $[1.6\,,\,3.9]$ & $[1.6\,,\,3.9]$ & $[1.6\,,\,3.9]$ & $[1.6\,,\,3.9]$ \\
$ n_\mathrm{s}  $ & $[0.8\,,\, 1.2]$ & -- & -- & -- & -- \\ 
$ \mathcal{N}  $ & -- & $[10\,,\,300]$ & $[10\,,\,300]$ & $[10\,,\,300]$ & $[10\,,\,300]$ \\
$\alpha$ & -- & -- & $[0\,,\,100]$ & $[0\,,\,1]$ & --\\

\hline \hline
\end{tabular} }
\end{center}
\caption{\small List of uniform prior distributions for cosmological parameters adopted for the different models. When considering the combinations of data involving the ACTPol CMB and lensing measurements, we introduce an additional Gaussian prior of $\tau=0.065\pm0.015$ for the optical depth at reionization.} 
\label{tab.priors}
\end{table}

To analyze the four potentials studied in this work and explore the observational constraints achievable with current CMB and large-scale structure probes, we follow this latter approach. We adopt the conventional power-law parametrizations (\autoref{PLS} and \autoref{PLT} for the scalar and tensor spectrum, respectively) and, for each model, we establish relationships among the inflationary parameters ($n_{s}$, $\alpha_{s}$, and $r$) by imposing the corresponding relations derived in \autoref{sec-inflation-models}. Specifically, we use \autoref{eq.Starob} for the Starobinsky model, \autoref{eq.alphaAtt} for $\alpha$-Attractors, \autoref{eq.PolInf} for polynomial inflation, and \autoref{eq.SUSY} for the SUSY potential. By assuming these relations, all inflationary parameters become interconnected, and be computed as functions of $\mathcal{N}$ and any other free parameters of the potential that will be varied in our exploration of the parameter space.

We compute the theoretical model using the Boltzmann integrator code \texttt{CAMB}~\cite{Lewis:1999bs,Howlett:2012mh} while in order to explore the parameter space of our models, we employ the publicly available sampler \texttt{COBAYA}~\cite{Torrado:2020xyz}. The code explores the posterior distributions of a given parameter space using the Monte Carlo Markov Chain (MCMC) sampler developed for \texttt{CosmoMC}~\cite{Lewis:2002ah} and tailored for parameter spaces with speed hierarchy implementing the ``fast dragging'' procedure developed in~\cite{Neal:2005}. Together with the usual $\Lambda$CDM parameters, (\textit{i.e.,} $A_\mathrm{s}$, $\Omega_{\rm b}h^2$,$\Omega_{\rm c}h^2$, $\theta_{\rm{MC}}$ and $\tau$), for the different inflationary models we sample over the number of last $e$-folds $\mathcal{N}$ between the horizon crossing and the end of inflation $\mathcal{N}$. For the $\alpha$-Attractor and Polynomial inflation we also include the additional free degree of freedom denoted as $\alpha$ (where, for Polynomial inflation, $\alpha=2(1-1/n)$). In \autoref{tab.priors} we summarize the prior distributions for the all the sampled parameters involved in our analysis that are chosen to be uniform across the range of variation, with the only exception of the optical depth ($\tau$) for which the prior distribution is chosen accordingly to the CMB datasets. The convergence of the chains obtained with this procedure is tested using the Gelman-Rubin criterion~\cite{Gelman:1992zz} and we choose as a threshold for chain convergence of $R-1 \lesssim 0.02 $.   

Concerning CMB and large-scale structure probes, our baseline data-sets consist combined with:
\begin{itemize}

\item The full Planck 2018 temperature and polarization (TT TE EE) likelihoods~\cite{Planck:2019nip,Planck:2018vyg,Planck:2018nkj} in combination with the Planck 2018 lensing likelihood~\cite{Planck:2018lbu}, reconstructed from the temperature 4-point correlation function. We combine this dataset with the 2018 B-modes polarization likelihood from the BICEP/Keck Collaboration~\cite{BICEP:2021xfz} and the Baryon Acoustic Oscillations (BAO) and Redshift Space Distortions (RSD) measurements from BOSS DR12~\citep{BOSS:2012dmf} and eBOSS DR16~\cite{Dawson:2015wdb}. We denote the final combination as \textbf{\textit{Planck+BK18+BAO+RSD}}.

\item The Atacama Cosmology Telescope temperature and polarization (TT TE EE) ACTPol-DR4 likelihood~\citep{ACT:2020frw}, in combination with the recent ACTPol-DR6 lensing likelihood~\cite{ACT:2023kun}. In this case we also assume a conservative Gaussian prior $\tau=0.065\pm0.015$. As for Planck, we combine this dataset with the 2018 B-modes polarization likelihood from the BICEP/Keck Collaboration~\cite{BICEP:2021xfz} and the Baryon Acoustic Oscillations (BAO) and Redshift Space Distortions (RSD) measurements from BOSS DR12~\citep{BOSS:2012dmf} and eBOSS DR16~\cite{Dawson:2015wdb}.  Similarly to the previous case, we denote the final combination as \textbf{\textit{ACTPol+BK18+BAO+RSD}}.

\end{itemize}

Finally, to conduct a model comparison, we first calculate the Bayesian evidence of each model and then estimate the corresponding Bayes factors (normalized for each dataset to the model which provides the best evidence). To do so, we utilize the publicly available package \texttt{MCEvidence}~\cite{Heavens:2017hkr,Heavens:2017afc}\footnote{The package is accessible at \href{https://github.com/yabebalFantaye/MCEvidence}{github.com/yabebalFantaye/MCEvidence}.}, that has been suitably modified to be compatible with \texttt{COBAYA}. To determine the level of preference for the optimal model, we exploit a revised version of the Jeffreys' scale derived from Ref.~\cite{Trotta:2008qt}. In particular, we consider the evidence to be inconclusive if $| \ln B_{ij}|  < 0.1$, weak if $0.1<| \ln B_{ij}|  < 1$,  moderate if $1<| \ln B_{ij}|  < 2.5$, and strong if $2.5<| \ln B_{ij}|  < 5$.

\begin{table*}
\begin{center}
\renewcommand{\arraystretch}{1.5}
\resizebox{\textwidth}{!}{
\begin{tabular}{l c c c c c c c c c c c c c c c }
\hline
\textbf{Parameter} & \textbf{ $\Lambda$CDM } & \textbf{ Starobinsky } & \textbf{$\alpha$ - Attractor } & \textbf{Polynomial inflation} & \textbf{SUSY} \\ 
\hline\hline

$ \Omega_\mathrm{b} h^2  $ & $  0.02245\pm 0.00013 $ & $  0.02246\pm 0.00013 $ & $  0.02248\pm 0.00013 $ & $  0.02261\pm 0.00014 $ & $  0.02246\pm 0.00013 $ \\ 
$ \Omega_\mathrm{c} h^2  $ & $  0.11908\pm 0.00090 $ & $  0.11894\pm 0.00090 $ & $  0.11873\pm 0.00090 $ & $  0.11682\pm 0.00084 $ & $  0.11897\pm 0.00091 $ \\ 
$ 100\theta_\mathrm{MC}  $ & $  1.04104\pm 0.00029 $ & $  1.04105\pm 0.00029 $ & $  1.04108\pm 0.00029 $ & $  1.04127\pm 0.00029 $ & $  1.04106\pm 0.00029 $ \\ 
$ \tau  $ & $  0.0593\pm 0.0073 $ & $  0.0607\pm 0.0066 $ & $  0.0607\pm 0.0066 $ & $  0.0663\pm 0.0073 $ & $  0.0605\pm 0.0067 $ \\ 
$ H_0  $ & $  67.78\pm 0.41 $ & $  67.85\pm 0.41 $ & $  67.95\pm 0.41 $ & $  68.80\pm 0.39 $ & $  67.84\pm 0.41 $ \\ 
$ \log(10^{10} A_\mathrm{s})  $ & $  3.054\pm 0.014 $ & $  3.056\pm 0.013 $ & $  3.055\pm 0.013 $ & $  3.060\pm 0.014 $ & $  3.056\pm 0.013 $ \\
$ n_\mathrm{s}  $ & $  0.9675\pm 0.0037 $ & $  0.9682\pm 0.0037 $ & $  0.9696\pm 0.0037 $ & $  0.9826\pm 0.0031 $ & $  0.9683\pm 0.0037 $ \\ 
$ \mathcal{N}  $ & -- & $  64^{+8}_{-9}\, ( 64^{+20}_{-20} ) $ & $  67^{+8}_{-10}\, ( 67^{+20}_{-20} ) $ & $  127\pm 30\, ( 127^{+70}_{-60} ) $ & $  64^{+8}_{-9}\, ( 64^{+20}_{-20} ) $ \\
$\alpha$ & -- & -- & $ < 7.44 \, (< 14.7)$ & $< 0.287\, (< 0.645)$ & --\\
\hline
$ r  $ & -- & $  0.00307\pm 0.00071\, ( 0.0031^{+0.0016}_{-0.0015} ) $ & $  0.016^{+0.010}_{-0.012}\, ( 0.016^{+0.023}_{-0.021} ) $ & $  0.074\pm 0.013\, ( 0.074^{+0.025}_{-0.025} ) $ & $  0.00204\pm 0.00047\, ( 0.00204^{+0.0010}_{-0.00098} ) $ \\ 
$ \alpha_s  $ & -- & $  -0.00051\pm 0.00012 $ & $  -0.00047\pm 0.00011 $ & $  \left(\,-14.8^{+6.3}_{-5.4}\,\right)\cdot 10^{-5} $ & $  -0.00051\pm 0.00012 $ \\ 
\hline
$\Delta \chi ^2$ & 0 & $-2.38$ & $-2.58$ & $26.0$ & $-1.48$\\
$\ln B_{ij} $ & -- & $0.11$ & $0$ & $14.34$ & $0.64$\\

\hline \hline
\end{tabular} }
\end{center}
\caption{\small The 68\% and 95\% CL constraints on various free and derived parameters obtained in a variety of inflationary potentials for the combined dataset \textit{Planck+BK18+BAO+RSD}. The horizontal line divides the constraints on the parameters from the corresponding predictions for $\alpha_s$ and $r$. Additionally we present the 
$\Delta \chi^2$  defined as $\Delta \chi^2 =$ ~$\chi^2$ (for the inflationary model)~$-$~$\chi^2$ (for $\Lambda$CDM) and the logarithm of the Bayes factors $\ln B_{ij}$.} 
\label{tab.results.P18}
\end{table*}

\begin{table}
\begin{center}
\renewcommand{\arraystretch}{1.5}
\resizebox{\textwidth}{!}{
\begin{tabular}{l c c c c c c c c c c c c c c c }
\hline

\textbf{Parameter} & \textbf{ $\Lambda$CDM } & \textbf{ Starobinsky } & \textbf{$\alpha$ - Attractor } & \textbf{Polynomial inflation} & \textbf{SUSY} \\ 
\hline\hline
$ \Omega_\mathrm{b} h^2  $ & $  0.02164\pm 0.00029 $ & $  0.02176\pm 0.00026 $ & $  0.02174\pm 0.00026 $ & $  0.02176\pm 0.00025 $ & $  0.02175\pm 0.00026 $ \\ 
$ \Omega_\mathrm{c} h^2  $ & $  0.1180\pm 0.0014 $ & $  0.1187\pm 0.0012 $ & $  0.1186\pm 0.0012 $ & $  0.1186\pm 0.0012 $ & $  0.1187\pm 0.0012 $ \\ 
$ 100\theta_\mathrm{MC}  $ & $  1.04214\pm 0.00062 $ & $  1.04216\pm 0.00062 $ & $  1.04214\pm 0.00063 $ & $  1.04213\pm 0.00061 $ & $  1.04215\pm 0.00062 $ \\ 
$ \tau  $ & $  0.083\pm 0.016 $ & $  0.071\pm 0.011 $ & $  0.071\pm 0.010 $ & $  0.071\pm 0.010 $ & $  0.071\pm 0.011 $ \\ 
$ H_0  $ & $  67.87\pm 0.57 $ & $  67.74\pm 0.52 $ & $  67.74\pm 0.52 $ & $  67.76\pm 0.50 $ & $  67.73\pm 0.52 $ \\ 
$ \log(10^{10} A_\mathrm{s})  $ & $  3.094\pm 0.027 $ & $  3.080\pm 0.020 $ & $  3.078\pm 0.020 $ & $  3.076\pm 0.019 $ & $  3.080\pm 0.020 $ \\

$ n_\mathrm{s}  $ & $  0.9999\pm 0.012 $ & $  0.9889^{+0.0056}_{-0.0034} $ & $  0.9904^{+0.0038}_{-0.0023} $ & $  0.9917^{+0.0014}_{-0.0010} $ & $  0.9890^{+0.0057}_{-0.0033} $ \\ 
$ \mathcal{N}  $ & -- & $ > 168\, (> 100 ) $ & $ > 202\, (> 127 ) $ & $ > 253\, (> 206 ) $ & $ > 172\, (> 100 ) $\\
$\alpha$ & -- & -- & unconstrained & $< 0.320\, (< 0.652)$ & --\\
\hline
$ r  $ & -- &  $0.00043^{+0.00025}_{-0.00048}\, (< 0.00120 ) $ & $  0.0114^{+0.0071}_{-0.0085}\, (<0.030) $ & $  0.0357^{+0.0055}_{-0.0072}\, ( 0.036^{+0.014}_{-0.013} ) $ & $  0.00028^{+0.00016}_{-0.00033}\, (< 0.000799 ) $ \\ 
$ \alpha_s  $ & -- & $  \left(\,-7.1^{+8.0}_{-4.1}\,\right)\cdot 10^{-5} $ & $  \left(\,-5.1^{+4.7}_{-2.6}\,\right)\cdot 10^{-5} $ & $  \left(\,-3.26^{+1.1}_{-0.75}\,\right)\cdot 10^{-5} $ & $  \left(\,-7.0^{+8.3}_{-4.1}\,\right)\cdot 10^{-5} $ \\ 
\hline
$\Delta \chi ^2$ & 0 & $0.26$ & $0.20$ & $2.04$ & $0.74$\\
$\ln B_{ij}$ & -- & $3.13$ & $0$ & $7.78$ & $2.78$\\

\hline \hline
\end{tabular} }
\end{center}
\caption{\small The 68\% and 95\% CL constraints on various free and derived parameters obtained from a variety of inflationary potentials for the combined dataset \textit{ACTPol+BK18+BAO+RSD}. The horizontal line divides the constraints on the parameters from the corresponding predictions for $\alpha_s$ and $r$. Additionally we present the 
$\Delta \chi^2$  defined as $\Delta \chi^2 =$ ~$\chi^2$ (for the inflationary model)~$-$~$\chi^2$ (for $\Lambda$CDM) and the logarithm of the Bayes factors $\ln B_{ij}$.} 
\label{tab.results.ACT}
\end{table}

\section{Results}
\label{sec-results}

In \autoref{tab.results.P18} and \autoref{tab.results.ACT}, we present the observational constraints on various inflationary potentials described in \autoref{sec-inflation-models} for the combination of datasets involving the Planck and the Atacama Cosmology Telescope measurements, respectively.

It is worth noting again that when assuming an inflationary potential, parameters such as $n_s$, $r$, and $\alpha_s$ are derived through consistency relations that vary from case to case\footnote{Notice that in our analysis of the inflationary potentials, we always include the running of the spectral index in the parameterization of the scalar spectrum. We tested for a few models (Starobinsky and $\alpha$-Attractor) that including or excluding the running does not lead to any dramatic differences in the constraints on the last $e$-folds $\mathcal{N}$. Thus, it does not alter the main conclusions derived in this work. This is primarily attributed to the fact that in these models the running is theoretically constrained to be very small, $\alpha_s \propto (1-n_s)^2$, thus producing subdominant effects on the spectrum. Despite this, we present the results including the running, as they provide useful predictions of its expected magnitude.}. Consequently, different inflationary models yield distinct predictions. Therefore, in \autoref{tab.results.P18} and \autoref{tab.results.ACT}, we directly compare our model-dependent results with those obtained for the same combination of data within a baseline $\Lambda$CDM model of cosmology where the inflationary parameters are constrained independently of  each other, which amounts in remaining agnostic about the inflationary  potential.
This comparison is useful to understand how adopting a specific inflationary theory alters the fit to data; how the discrepancies between data-sets affect the results for model selection; and eventually identifying a model or a set of models that broadly align with all these CMB observations spanning different angular scales. Let us also point out that, from the constraints obtained for the various models in the following subsections, we can derive any other quantity of interest for the models or potentials and have complete information. Interested readers can find an example in \autoref{appendix}, where we study the implications of
the discussed constraints for the determination of the energy scale that suppresses modifications of gravity in the Starobinsky model.

\subsection{Planck+BK18+BAO+RSD}

We start focusing on the analysis conducted using the dataset including the Planck CMB measurements, as summarized in \autoref{tab.results.P18}. It becomes evident that, apart from the Polynomial inflation case, the Starobinsky model, $\alpha$-Attractor, and SUSY, exhibit similar behavior in terms of the cosmological parameters. For example, the spectral index of the primordial scalar spectrum $n_s$ obtained in the Starobinsky model, $\alpha$-Attractors, and SUSY are in perfect agreement with the results derived assuming a baseline $\Lambda$CDM model and remaining blind on the specific model of Inflation. On the other hand, for the Polynomial inflation model, the spectral index $n_s$ ($= 0.9826\pm 0.0031$ at 68\% CL) is sensibly higher compared to $\Lambda$CDM.

The consistency relations among inflationary parameters allow us to derive accurate \textit{predictions}, specific to each model, regarding additional quantities listed in \autoref{tab.results.P18}, such as the running of the spectral index and tensor amplitude. It is important to emphasize that these results should be considered as specific model-dependent predictions, rather than actual detections\footnote{In other words, the interpretation of the results provided for $\alpha_s$ and $r$ is as follows: given the constraints on the other measured inflationary parameters, we expect that, within each specific model, the tensor amplitude and the running of the spectral index will fall within the reported ranges at a 68\% or 95\% CL.}. As concerns the predictions for $\alpha_s$ in these inflationary models, we have at 68\% CL  $\alpha _s = -0.00051\pm 0.00012 $ (Starobinsky model), $ \alpha _s =  -0.00047\pm 0.00011$ ($\alpha$-Attractor), and $\alpha _s = -0.00051\pm 0.00012$ (SUSY), i.e., negatives values ruling out the null value at more than $3\sigma$. Similarly, in the case of Polynomial inflation, there is still a prediction for the running of the scalar spectral index not zero, even if an order of magnitude smaller than the previous three cases. 
Comparing these results with those reported in Table 3 of~\cite{Planck:2018jri}, where instead $\alpha_s$ is treated as an independent free parameter, reveals a notable decrease in the average values of $\alpha_s$ along with its uncertainties. The same consideration applies to the tensor-to-scalar ratio, $r$, where our findings suggest it to be significantly smaller than the values obtained from Planck 2018~\cite{Planck:2018jri} for Starobinsky, $\alpha$-Attractor, and SUSY while in Polynomial inflation it is found one order of magnitude higher than in the other inflationary scenarios, yet it is consistently predicted to be non-zero for all the inflationary models. All these results present testable predictions for the upcoming stage-4 CMB experiments~\cite{BICEP3,CLASS,LBIRD,Abazajian:2019eic,CMB-S4,SimonsObservatory:2018koc,NASAPICO:2019thw,CMB-HD:2022bsz}, which are anticipated to achieve a sensitivity on the amplitude of primordial tensor perturbations and the running of the spectral index comparable to our findings. 

As concerns the additional free parameters of the different inflationary potentials, it is important to note that for this particular dataset, the parameter $\alpha$ in the $\alpha$-Attractor model can be constrained to $\alpha<14.7$ while for Polynomial inflation the additional free parameter $\alpha$ has an upper limit of $\alpha < 0.645$ (both at 95\% CL).

When considering the number of last $e$-folds of expansion, the constraints on various inflationary potentials indicate similar values of $\mathcal{N}$ in the range of $50$ to $70$ for models such as the Starobinsky, $\alpha$-Attractor, and SUSY inflation. This range of $e$-folds is in agreement with our expectations for good inflationary models that can generate a sufficiently long period of exponential almost-de Sitter expansion able to account for the observed large-scale flatness and homogeneity in the Universe~\cite{Lidsey:1995np,Liddle:2003as,deHaro:2023xcc}. However, it is important for this range not to be excessively long in order to remain also consistent with the upper bounds derived in the literature, typically ranging between 60 and 70 $e$-folds of expansion\footnote{Notice that these bounds are based on the relationship between the number of $e$-folds $\mathcal N$ and other inflationary observables, such as the reheating temperature and the energy scale of inflation. For more details, refer to \autoref{eq:N} or Refs.~\cite{Lidsey:1995np,Liddle:2003as,deHaro:2023xcc}. It is also worth noting that it is theoretically possible to achieve a greater number of $e$-folds of expansion by reducing the energy density during inflation, thereby significantly lowering the energy scale at the end of inflation. However, such an evolution is not viable on observable scales where perturbations are observed because it would lead to a spectrum that deviates significantly from scale-invariance. Nevertheless, in principle, it could occur at the later stages, resulting in an extreme upper bound of $\mathcal N \lesssim 100$, see Ref.~\cite{Liddle:2003as}.}~\cite{Liddle:2003as}. Considering the polynomial inflation potential, the number of last $e$-folds is found to be $\mathcal{N} = 127\pm 30$ at a 68\% CL ($\mathcal{N} = 132^{+70}_{-60}$ at a 95\% CL).  These result suggests that we are approaching a range of $e$-folds that might exceed the upper limit. However, due to the large error-bars, we cannot definitively dismiss this potential, as it still falls within an acceptable range at the 95\% CL.

Finally, when considering a specific model for the early Universe, the subsequent cosmological evolution is determined by the initial conditions set by the inflationary potential. Interestingly, various studies have suggested that the inflationary sector of the theory may play a significant role in addressing (part of) the tension between the expansion rate of the Universe measured through direct local distance ladder measurements and the value inferred from CMB observations~\cite{DiValentino:2018zjj,Ye:2022efx,Jiang:2022uyg,Jiang:2022qlj,Takahashi:2021bti,Lin:2022gbl,Hazra:2022rdl,Braglia:2021sun,Keeley:2020rmo,Jiang:2023bsz,Peng:2023bik}. Therefore it is worth examining the estimated values of the Hubble constant, $H_0$, in these four models. In the Starobinsky model, $\alpha$-Attractor, and SUSY, we observe that they closely resemble the values obtained from Planck~\cite{Planck:2018vyg} assuming a $\Lambda$CDM scenario. Consequently, the $5\sigma$ tension between these values and the measurements from the SH0ES collaboration~\cite{Riess:2021jrx} remains unresolved~\cite{DiValentino:2021izs,Perivolaropoulos:2021jda,Schoneberg:2021qvd,Abdalla:2022yfr}.  Interestingly, in the Polynomial inflation case, the Hubble constant assumes a slightly higher value compared to the Planck 2018 value for a $\Lambda$CDM model, thus mildly alleviating the Hubble tension at a significance level of $3.9\sigma$, further emphasizing the importance of determining precise predictions for the inflationary potential when evaluating current cosmological and astrophysical observations. 

Moving forward, our focus turns to model selection. The main goal we would like to address is to assess which inflationary model(s) can offer a more consistent explanation of the observed CMB data released by the Planck collaboration and understand to what extent these interpretations agree with measurements obtained from various CMB and large scale structure surveys. To do so we compare the results obtained for the different potentials. An inspection of \autoref{tab.results.P18} reveals that the inclusion of Starobinsky inflation leads to a modest improvement of $\Delta\chi^2 = -2.38$ in fitting the combined Planck dataset, when compared to the baseline $\Lambda$CDM cosmology. However, in the model comparison undertaken in this study, Starobinsky inflation is not identified as the best-performing model for Planck. Notably, the adoption of an $\alpha$-Attractor potential yields a slightly larger improvement of $\Delta\chi^2 = -2.58$, resulting as the most favorable model also in the Bayesian model comparison. The SUSY potential also exhibits good performance, with an improvement in the fit of $\Delta\chi^2 = -1.48$ compared to $\Lambda$CDM. On the other hand, the main difference of the Polynomial inflation from the other three potentials is evident in the $\Delta \chi^2$ ($= 26.0$) which takes the larger value for this model. This overall deterioration in the $\chi^2$ is further emphasized during the model comparison, as strong evidence emerges indicating that the Polynomial inflation model is highly disfavored when compared to the other models.

\subsection{ACTPol+BK18+BAO+RSD}
We now shift our focus to the analysis of the same models but for the dataset combination involving the ACTPol measurements of the CMB at small scales. The results are shown in \autoref{tab.results.ACT}. 
It is worth noting that the replacement of CMB data from Planck 2018 with ACTPol is a crucial step to study the very same inflationary potential under different angular scales in the cosmic microwave background. Indeed, comparing the results from these two independent probes, we can assess their consistency and understand the impact of the differences in the most recent CMB observations on model selection.
 
Replacing Planck with ACTPol (DR4+DR6) leads to distinct results for the inflationary parameters. Indeed, this dataset combination predicts a higher spectral index, $n_s$, compared to Planck 2018 results, as is already the case for the $\Lambda$CDM model, see also Refs.~\cite{ACT:2020gnv,Giare:2022rvg}. Moreover, in most models, the indication for a non-zero running of the spectral index is reduced to slightly more than $1\sigma$. The tensor-to-scalar ratio exhibits a further reduction when using the ACTPol dataset combination compared to Planck. Specifically, we predict mean values that are up to one order of magnitude lower for the Starobinsky and SUSY cases and a factor of 2 for the Polynomial inflation scenario. Regarding the free parameters of the inflationary potentials, the $\alpha$ in the $\alpha$-Attractor case remains unconstrained despite the large prior used in the MCMC analysis (refer to \autoref{tab.priors}). This lack of constraint can be attributed to the following: in this model, the quantity $r$ is proportional to $\alpha / \mathcal{N}^2$. When replacing Planck with ACTpol, only a lower limit can be obtained on the number of $e$-folds between the horizon exit and the end of inflation. Consequently, a strong degeneracy arises between $\mathcal{N}$ and $\alpha$: in the absence of a two-tail limit on the former, higher values of $\alpha$ can be always compensated by increasing $\mathcal{N}$. 
This prevents us from achieving an upper
bound, unlike with the previous combination of data. 
Conversely, for Polynomial inflation we do not find any significant changes in terms of the constraints on $\alpha$ whose upper limit reads $\alpha < 0.652$ (at a 95\% CL). As concerns the values obtained for the Hubble constant in all four models are compatible with the Planck value within the $\Lambda$CDM paradigm~\cite{Planck:2018vyg}. Consequently, they are in disagreement with the measurement from the SH0ES collaboration~\cite{Riess:2021jrx}.

However, a notable discrepancy arises when comparing the number of last $e$-folds before the end of inflation obtained by ACTPol with those derived by the Planck measurements at larger angular scale. Examining \autoref{tab.results.ACT}, for ACTPol we find the following values for the number of $e$-folds at 68\% (95\%) CL:
\begin{itemize}[leftmargin= *]
\item $\mathcal{N} > 168$ ($\mathcal{N} > 100$) for the Starobinsky model;
\item $\mathcal{N} > 202$ ($\mathcal{N} > 127$) for $\alpha$-Attractor models;
\item $\mathcal{N} > 253$ ($\mathcal{N} > 206$)  for Polynomial inflation;
\item $\mathcal{N} > 172$ ($\mathcal{N} > 100$) for the SUSY potential.
\end{itemize}

\begin{figure}[t]
    \centering
   \includegraphics[width=0.6\textwidth]{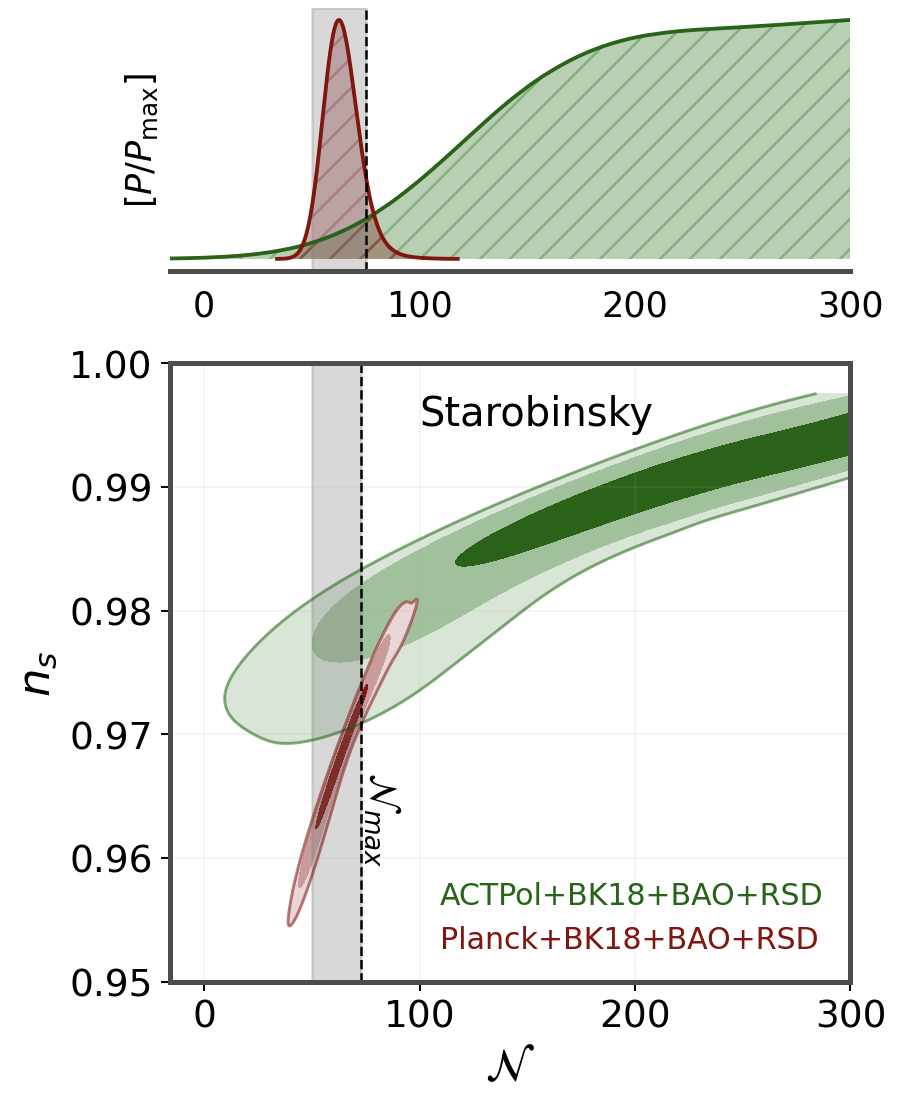}
   \caption{\small 2D contours at 68\%, 95\%, and 99\% CL and 1D posteriors in the ($n_s$, $\mathcal{N}$) plane for the Starobinsky model. The grey vertical band refers to the typical range of folds expansion $\mathcal{N} \in [50, \mathcal{N}_{max}]$, expected in standard inflation. The upper limit, $\mathcal{N}_{max} \le 75$, estimated by \autoref{eq.Nmax}, is represented by the black dashed line. Very similar results are obtained for all the potentials analyzed in this study.}
    \label{fig:starobinsky}
\end{figure}

These very large lower limits for the $e$-folds of expansion reflect the preference of ACTPol data for higher values of the spectral index of inflationary perturbations: as $n_s$ is pushed to higher values in the direction of unity, the number of $e$-folds also increases as clearly shown in \autoref{fig:starobinsky} in the case of the Starobinsky model. However, such large values of $\mathcal{N}$ raise concerns for different inflationary mechanisms, spanning from the most common Starobisnky and $\alpha$-attractor potentials to quintessential models of inflation. For instance, it is well-known that the number of last $e$-folds in standard inflation is smaller than in Quintessential Inflation. As a matter of fact, the latter features a kination phase between the end of inflation and the onset
of radiation domination. During such a phase, the effective equation-of-state parameter of the cosmological fluid $w$ is equal to one\footnote{Notice that in Quintessential Inflation, there is a short transition period from the end of inflation ($w=-1/3$) to the beginning of kination phase ($w=1$). While this transition is dependent on the specific potential, it has minimal impact on the number of $e$-folds (typically no more than 1 $e$-fold) and we can confidently assume $w=1$.}. As a consequence, the number of last $e$-folds can be related to the reheating temperature ($T_{\rm reh}$) as \cite{deHaro:2023xcc}:


\begin{equation}
\mathcal{N} \approx 54.47 + \frac{1}{2}\ln \epsilon_* + \frac{1}{3}\ln \left(\frac{M_{pl}^2}{T_{\rm{reh}}H_{\rm{end}}}\right),
\label{eq:N}
\end{equation}
where $H_{\rm{end}}$ represents the value of the Hubble rate at the end of inflation. To establish a reasonable upper limit on the number of folds applicable to the most typical single-field inflationary models, we made the robust assumption that $\epsilon_{\star}\lesssim 10^{-2}$. This limit can be considered very conservative given the current constraint on the tensor amplitude $r=16\epsilon_{\star}<0.035$~\cite{BICEP:2021xfz}\footnote{We should note that a stricter upper limit on $\epsilon$ would result in tighter upper bound on $\mathcal{N}$.}. Regarding $H_{\text{end}}$, we previously noticed that extremely low values could lead to significantly higher numbers of $e$-folds. However, the vast majority of inflationary models predict an inflationary energy scale $H_{\text{end}}\sim 10^{-6} M_{\text{pl}}$. Therefore, also in this case we maintain a conservative approach by assuming $H_{\text{end}}\gtrsim 10^{-8} M_{\text{pl}}$, thus leaving aside the exploration of controversial scenarios with a significant evolution of the inflationary energy density, typically hard to realize within single field potentials such as those of analyzed in this work. Under this working assumptions, we can obtain: 
\begin{equation}
\mathcal{N} \leq  58.3 + \frac{1}{3}\ln \left(\frac{M_{pl}}{T_{\rm{reh}}}\right).
\end{equation}
$\mathcal{N}$ reaches its maximum value when considering the minimum viable reheating temperature of $T_{\rm{reh}} \sim 1$ MeV ($\sim 5 \times 10^{-22}M_{pl}$), resulting in the upper bound:
\begin{equation}
\mathcal{N}_{max} < 75
\label{eq.Nmax}
\end{equation}
which is very consistent with previous findings in the literature~\cite{Lidsey:1995np,Liddle:2003as,deHaro:2023xcc}. On the other hand, for the Starobinsky, $\alpha$-Attractor, and exponential SUSY models, we have observed that the number of $e$-folds can be approximated as ${\mathcal N} \approx \frac{2}{1-n_s}$. Thus, the minimum value of the spectral index corresponds to the minimum value of $e$-folds for these three potential models.
Using the ACTPol's results at the $2\sigma$ level from \autoref{tab.results.ACT}, we find:
\begin{equation}
\mathcal{N}_{min} \gtrsim 100,
\end{equation}
which clearly indicates that these inflationary models are all disregarded by the ACTPol data at $\sim3\sigma$, as shown in \autoref{fig:starobinsky} for the Starobinsky case. 

This result may come as a very surprising finding since these models of inflation are typically regarded as well-established benchmark scenarios for testing the inflationary paradigm in the first place. Consequently, most of the current planned next-generation CMB experiments are specifically designed to probe their predictions. However, the general inadequacy of these models to provide a satisfactory fit to the ACTPol measurements of the damping tail is also reflected in the model comparison. In the case of Starobinsky inflation, the fit to the ACTPol dataset worsens with an increase in $\Delta\chi^2=0.26$ compared to $\Lambda$CDM. Similar results are observed for all the other models: $\alpha$-Attractors ($\Delta\chi^2=0.20$), Polynomial inflation ($\Delta \chi^2=2.04$) and the SUSY potential ($\Delta\chi^2=0.74$). In this model comparison, the $\alpha$-Attractor potential emerges again as
the best-performing option, despite its significant $\sim 3 \sigma$ tension with regard to the number of last $e$-folds. 

\section{Summary and Conclusions}
\label{sec-conclu}

The theory of inflation has proven to be highly successful in describing early Universe physics. In this regard, precise observations of the CMB from missions like WMAP, Planck and most recently ACTPol, have played a crucial role, providing stringent constraints on inflationary models which do not always agree. Therefore, given the absence of detection of B-mode polarization and the emerging differences among independent CMB observations, identifying a universally preferred model (or group of models) based on all available CMB data remains an open issue and a large number of competing inflationary potentials can be regarded as equally plausible based on current records. 

In this paper, we further explore this difficulty. As a case study, we examine four well-known inflationary potentials using independent CMB and lensing data at different angular scales released by Planck and ACTPol, in combination with the 2018 B-modes polarization likelihood from the BICEP/Keck Collaboration and the Baryon Acoustic Oscillations (BAO) and Redshift Space Distortions (RSD) measurements from BOSS DR12 and eBOSS DR16. The constraints on the inflationary models are summarized in \autoref{tab.results.P18} and \autoref{tab.results.ACT}, for the datasets involving Planck and ACTPol, respectively. 

Our results clearly show that the discrepancy between the CMB observations at larger angular scales from Planck and the highly precise measurements of the damping tail provided by ACTPol leads to significant differences when the same inflationary potentials are assumed in the cosmological model and analyzed. These discrepancies have a substantial impact on the predictions of various models for essential quantities such as the tensor amplitude and the running of the scalar spectral index, whose values dictate the theoretical, phenomenological, and experimental perspectives of the field. Moreover, our analysis proves that these differences also yield discordant conclusions regarding the preferred model. In this regard, a significant finding is that while commonly known inflationary potentials such as Starobinsky and $\alpha$-attractor inflation are consistent with Planck and BICEP/Keck temperature and polarization data, they fail to adequately explain the small-scale CMB observations provided by ACTPol. It is well known that this dataset indicates a preference for higher values of the spectral index of primordial inflationary perturbations, which would require a significantly longer number of $e$-folds between horizon exit and the end of inflation. As a result, if we set aside observational systematics and consider the differences between ACTPol and the Planck satellite as genuine, inflationary potentials where $n_s\propto 1-{n}/\mathcal{N}$  with $n\geq 2$,  become inadequate in addressing these measurements, contributing to the existing uncertainty in the field. Potential solutions to address this issue could involve considering more elaborate inflationary models as well as making adjustments to the broader cosmological framework.

Regarding the first possibility, a few intriguing avenues have recently emerged in the literature. While not aiming for exhaustive coverage, we spotlight models similar to the $\alpha$-attractor generalization of the hybrid inflation scenario introduced in Ref.~\cite{Kallosh:2022ggf}. In this model, the predictions for the inflationary observables are gauged by an additional uplift parameter, and, based on its value, the model’s predictions can gradually shift from the most typical Starobinsky and $\alpha$-attractor scenarios preferred by Planck to the higher values of $n_s$ favored by ACTPol, eventually converging to a secondary attractor point at $n_s = 1$ in the large uplift limit. Notably, these models are suitable for describing both regimes and can accommodate values of $n_s \simeq 1$ without exceeding $\mathcal N \sim 50 - 60$ $e$-folds of expansion. However, further analyses are needed to assess their ability to reconcile the differences between experiments, especially when they are assumed from the outset of the analysis. Alternatively, given the different scales probed by the two experiments, the discrepancy in the value of the spectral index might be alleviated by inflationary models with satiable scale-dependence of the primordial spectrum. For instance, in Ref.~\cite{Giare:2022rvg} it was argued that a positive running $\alpha_s$ could help mitigate the differences between the two probes. 

In a broader context, while it remains plausible that the disparity between two independent yet conflicting measurements of the same parameter could be mitigated within certain inflationary models, it is worth considering the possibility that these differences recast the presence of new physics manifesting itself at different scales. The values of cosmological parameters inferred from CMB observations rely on the underlying cosmological model assumed in the analysis, and $n_s$ is no exception. Therefore, we cannot disregard the possibility that the tension is simply rooted in the limitations of the standard cosmological model and an argument supporting this avenue comes from Refs.~\cite{Giare:2022rvg,DiValentino:2022rdg}, where it has been pointed out how the difference concerning the value of $n_s$ between Planck and ACTPolt is entirely absent in extended cosmological models with additional parameters. Hence, one could entail introducing new physics whose impact might counterbalance the disparities causing the shift in $n_s$ when a $\Lambda$CDM cosmology is assumed. To round out the picture, we shall mention two promising (albeit opposite) mechanisms that may help restore the agreement in the inflationary sector of the theory, while leaving more accurate analyses of these scenarios suitable for future studies. On one side, models where the radiation energy density in the early Universe is significantly lower than expected in the Standard Model of particle physics have proved effective in reducing the \textit{global} tension between the two experiments producing a shift in the value of $n_s$ measured by ACTPol and bringing it back in agreement with Planck~\cite{Giare:2022rvg,DiValentino:2022rdg}. On the other hand, as extensively documented in the literature~\cite{Calabrese:2011hg,Poulin:2018dzj,Niedermann:2019olb,Niedermann:2020dwg, Murgia:2020ryi, Herold:2021ksg, Reeves:2022aoi, Jiang:2022uyg,Niedermann:2023ssr,Cruz:2023lmn,Eskilt:2023nxm}, cosmological models featuring additional scalar degrees of freedom in the early Universe, such as Early Dark Energy~\cite{Poulin:2018cxd, Hill:2021yec,Poulin:2023lkg}, often exhibit a preference for a slightly larger spectral index, possibly shifting the value inferred by Planck in the direction favored by ACTPol (which in turn shows a $\sim 3\sigma$ preference for such scenarios~\cite{Hill:2021yec}).

Therefore, we conclude that, despite significant efforts to explore various inflationary scenarios, further investigations and refinements may be needed to achieve a more complete understanding of both the inflationary dynamics and the early Universe.

\acknowledgments
\noindent  
We thank the referee for several useful comments that have been helpful to increase the overall quality of the manuscript.   WG thanks Carsten van de Bruck for interesting discussions and suggestions around the subject of this work. 
SP acknowledges the financial support from  the Department of Science and Technology (DST), Govt. of India under the Scheme  ``Fund for Improvement of S\&T Infrastructure (FIST)'' [File No. SR/FST/MS-I/2019/41]. EDV is supported by a Royal Society Dorothy Hodgkin Research Fellowship.  WY was supported by the National Natural Science Foundation of China under Grants No. 12175096 and No. 11705079, and Liaoning Revitalization Talents Program under Grant no. XLYC1907098. 
JdH is supported by the Spanish grant 
PID2021-123903NB-I00
funded by MCIN/AEI/10.13039/501100011033 and by ``ERDF A way of making Europe''. AM thanks TASP, iniziativa specifica INFN for financial support.
This article is based upon work from COST Action CA21136 Addressing observational tensions in cosmology with systematics and fundamental physics (CosmoVerse) supported by COST (European Cooperation in Science and Technology).
We acknowledge IT Services at The University of Sheffield for the provision of services for High Performance Computing.

\appendix

\section{Implications for the inflationary potentials and fundamental physics}
\label{appendix}

\begin{figure}[t!]
    \centering
   \includegraphics[width=0.5\textwidth]{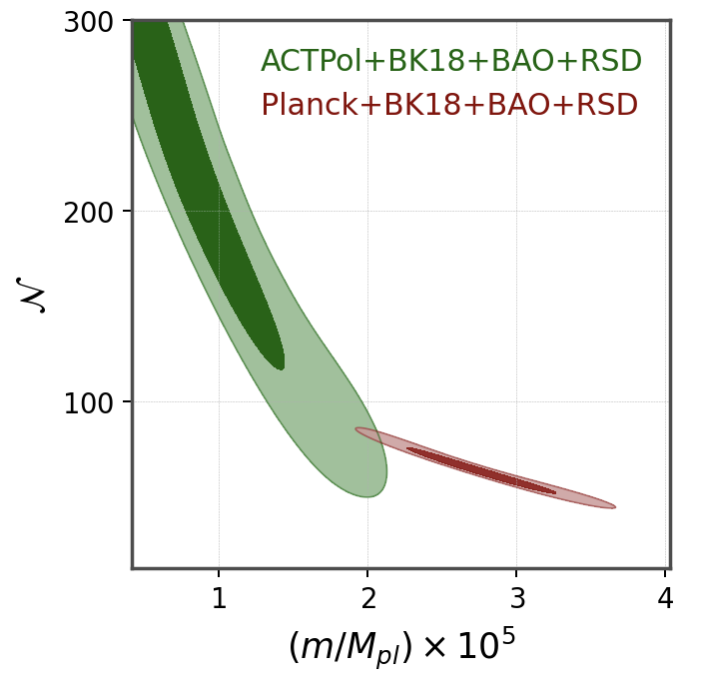}
   \caption{\small 2D contours at 68\% and 95\% CL in the plane ($m$, $\mathcal{N}$) where $m$ denotes the mass scale present in the modified $f(R)$-gravity relation $f(R) = R + R^2/m^2$. The mismatch in the number of last $e$-folds leads to different predictions for the energy scale associated with modified gravity corrections.}
    \label{fig:m_starobinsky}
\end{figure}

In this work, by considering well-established relationships among the number of last $e$-folds $\mathcal{N}$ and other observable quantities, such as the spectral index of scalar perturbations and the tensor-to-scalar ratio, we have demonstrated that CMB observations at different angular scales result in disparate predictions for benchmark inflationary models.

In this appendix, our primary goal is to demonstrate that all other parameters of the potential, not explicitly appearing in these equations, can still be derived when the mentioned quantities are constrained. Moreover, we will explore how these variations introduce uncertainty, especially in quantities related to fundamental physics. This additional layer of uncertainty extends beyond the model itself and influences its implications for fundamental physics.

As a case study, we consider again the Starobinsky model and derive constraints on the parameter $m$ that appears in the modified $f(R)$-gravity relation $f(R) = R + R^2/m^2$. This parameter holds particular significance as it represents the energy scale associated with corrections in modified gravity. Hence, any predictions for $m$ may serve as a compass to gauge the expected energy-scale of deviations from General Relativity.

To infer constrain on $m$, one can use the amplitude of the primordial spectrum of the scalar perturbations~\cite{Bassett:2005xm},
\begin{eqnarray}\label{spectrum}
A_s={\mathcal P}_{\zeta}(k_*)=\frac{H_*^2}{8\pi^2\epsilon_*M_{pl}^2}
\end{eqnarray}
Since at the horizon crossing for the Starobinsky model we have  $V_*\cong \frac{M_{pl}^2 m^2}{8} \Longrightarrow H_*^2=\frac{m^2}{24}$ and $\epsilon_*\cong \frac{3}{16}(1-n_s)^2$, the parameter $m$ can be expressed in terms of the aforementioned quantities as
\begin{equation}
\left(\frac{m}{M_{pl}}\right)^2 \simeq 36\pi^2A_s\left(1-n_s\right)^2
\end{equation}

Focusing on the analysis conducted using the dataset including the Planck CMB measurements, we obtain a 68\% CL constraint on $m$ which, in units of Planck mass, reads:
\begin{equation}
(m/M_{\text{pl}}) = \left(2.76\pm0.32\right) \times 10^{-5} \quad \text{for Planck+BK18+BAO+RSD}
\end{equation}
On the other hand, considering the dataset including the small-scale CMB measurements provided by ACT, the results for the same parameter are as follows:
\begin{equation}
(m/M_{\text{pl}}) = \left(0.98^{+0.29}_{-0.47}\right) \times 10^{-5} \quad \text{for ACTPol+BK18+BAO+RSD}
\end{equation}

These results demonstrate, first and foremost, that by following the approach adopted in this study, where inflation models are assumed from the outset in the analysis, all parameters of the model can be constrained by the data quite accurately. Additionally, it is worth noting the differences in the predicted energy scale that suppresses modified gravity terms by the two datasets. As observed from \autoref{fig:m_starobinsky}, such differences primarily arise from the mismatch in the number of $e$-folds predicted by the two datasets within this inflationary model, ultimately originating in the discrepant values of the spectral index as measured by the Planck satellite data and the Atacama Cosmology Telescope. This leads us to consider the importance of obtaining \textit{precise} predictions from cosmological observables to achieve reliable predictions for fundamental physics. Indeed, we would like to emphasize that even mild-to-moderate disagreements in the values of cosmological parameters may introduce some uncertainty when data are used for constraining fundamental physics.

These findings highlight the importance of accurately characterizing cosmological observables to improve the robustness and accuracy of our conclusions in the realm of fundamental physics.

\bibliographystyle{unsrt}
\bibliography{biblio}

\begin{thebibliography}{100}

\bibitem{Guth:1980zm}
Alan~H. Guth.
\newblock {The Inflationary Universe: A Possible Solution to the Horizon and
  Flatness Problems}.
\newblock {\em Phys. Rev. D}, 23:347--356, 1981.

\bibitem{Sato:1980yn}
K.~Sato.
\newblock {First Order Phase Transition of a Vacuum and Expansion of the
  Universe}.
\newblock {\em Mon. Not. Roy. Astron. Soc.}, 195:467--479, 1981.

\bibitem{Linde:1981mu}
Andrei~D. Linde.
\newblock {A New Inflationary Universe Scenario: A Possible Solution of the
  Horizon, Flatness, Homogeneity, Isotropy and Primordial Monopole Problems}.
\newblock {\em Phys. Lett. B}, 108:389--393, 1982.

\bibitem{Starobinsky:1980te}
Alexei~A. Starobinsky.
\newblock {A New Type of Isotropic Cosmological Models Without Singularity}.
\newblock {\em Adv. Ser. Astrophys. Cosmol.}, 3:130--133, 1987.

\bibitem{1980ApJ...239..428K}
E.~W. {Kolb} and S.~{Wolfram}.
\newblock {Spontaneous symmetry breaking and the expansion rate of the early
  universe}.
\newblock {\em APJ}, 239:428--432, July 1980.

\bibitem{Fixsen:1993rd}
D.~J. Fixsen et~al.
\newblock {Cosmic microwave background dipole spectrum measured by the COBE
  FIRAS}.
\newblock {\em Astrophys. J.}, 420:445, 1994.

\bibitem{Bennett:1996ce}
C.~L. Bennett, A.~Banday, K.~M. Gorski, G.~Hinshaw, P.~Jackson, P.~Keegstra,
  A.~Kogut, George~F. Smoot, D.~T. Wilkinson, and E.~L. Wright.
\newblock {Four year COBE DMR cosmic microwave background observations: Maps
  and basic results}.
\newblock {\em Astrophys. J. Lett.}, 464:L1--L4, 1996.

\bibitem{Hinshaw:2012aka}
G.~Hinshaw et~al.
\newblock {Nine-Year Wilkinson Microwave Anisotropy Probe (WMAP) Observations:
  Cosmological Parameter Results}.
\newblock {\em Astrophys. J. Suppl.}, 208:19, 2013.

\bibitem{WMAP:2012fli}
C.~L. Bennett et~al.
\newblock {Nine-Year Wilkinson Microwave Anisotropy Probe (WMAP) Observations:
  Final Maps and Results}.
\newblock {\em Astrophys. J. Suppl.}, 208:20, 2013.

\bibitem{WMAP:2003syu}
H.~V. Peiris et~al.
\newblock {First year Wilkinson Microwave Anisotropy Probe (WMAP) observations:
  Implications for inflation}.
\newblock {\em Astrophys. J. Suppl.}, 148:213--231, 2003.

\bibitem{Planck:2019nip}
N.~Aghanim et~al.
\newblock {Planck 2018 results. V. CMB power spectra and likelihoods}.
\newblock {\em Astron. Astrophys.}, 641:A5, 2020.

\bibitem{Planck:2018nkj}
N.~Aghanim et~al.
\newblock {Planck 2018 results. I. Overview and the cosmological legacy of
  Planck}.
\newblock {\em Astron. Astrophys.}, 641:A1, 2020.

\bibitem{Planck:2013jfk}
P.~A.~R. Ade et~al.
\newblock {Planck 2013 results. XXII. Constraints on inflation}.
\newblock {\em Astron. Astrophys.}, 571:A22, 2014.

\bibitem{Planck:2015sxf}
P.~A.~R. Ade et~al.
\newblock {Planck 2015 results. XX. Constraints on inflation}.
\newblock {\em Astron. Astrophys.}, 594:A20, 2016.

\bibitem{Planck:2018jri}
Y.~Akrami et~al.
\newblock {Planck 2018 results. X. Constraints on inflation}.
\newblock {\em Astron. Astrophys.}, 641:A10, 2020.

\bibitem{ACT:2020frw}
Steve~K. Choi et~al.
\newblock {The Atacama Cosmology Telescope: a measurement of the Cosmic
  Microwave Background power spectra at 98 and 150 GHz}.
\newblock {\em JCAP}, 12:045, 2020.

\bibitem{ACT:2020gnv}
Simone Aiola et~al.
\newblock {The Atacama Cosmology Telescope: DR4 Maps and Cosmological
  Parameters}.
\newblock {\em JCAP}, 12:047, 2020.

\bibitem{ACT:2023kun}
Mathew~S. Madhavacheril et~al.
\newblock {The Atacama Cosmology Telescope: DR6 Gravitational Lensing Map and
  Cosmological Parameters}.
\newblock 4 2023.

\bibitem{SPT-3G:2014dbx}
B.~A. Benson et~al.
\newblock {SPT-3G: A Next-Generation Cosmic Microwave Background Polarization
  Experiment on the South Pole Telescope}.
\newblock {\em Proc. SPIE Int. Soc. Opt. Eng.}, 9153:91531P, 2014.

\bibitem{SPT-3G:2021eoc}
D.~Dutcher et~al.
\newblock {Measurements of the E-mode polarization and temperature-E-mode
  correlation of the CMB from SPT-3G 2018 data}.
\newblock {\em Phys. Rev. D}, 104(2):022003, 2021.

\bibitem{SPT-3G:2022hvq}
L.~Balkenhol et~al.
\newblock {A Measurement of the CMB Temperature Power Spectrum and Constraints
  on Cosmology from the SPT-3G 2018 TT/TE/EE Data Set}.
\newblock 12 2022.

\bibitem{Martin:2013tda}
Jerome Martin, Christophe Ringeval, and Vincent Vennin.
\newblock {Encyclop\ae{}dia Inflationaris}.
\newblock {\em Phys. Dark Univ.}, 5-6:75--235, 2014.

\bibitem{Lyth:1998xn}
David~H. Lyth and Antonio Riotto.
\newblock {Particle physics models of inflation and the cosmological density
  perturbation}.
\newblock {\em Phys. Rept.}, 314:1--146, 1999.

\bibitem{Linde:2007fr}
Andrei~D. Linde.
\newblock {Inflationary Cosmology}.
\newblock {\em Lect. Notes Phys.}, 738:1--54, 2008.

\bibitem{Baumann:2014nda}
Daniel Baumann and Liam McAllister.
\newblock {\em {Inflation and String Theory}}.
\newblock Cambridge Monographs on Mathematical Physics. Cambridge University
  Press, 5 2015.

\bibitem{Leach:2002ar}
Samuel~M. Leach, Andrew~R. Liddle, Jerome Martin, and Dominik~J Schwarz.
\newblock {Cosmological parameter estimation and the inflationary cosmology}.
\newblock {\em Phys. Rev. D}, 66:023515, 2002.

\bibitem{Boubekeur:2005zm}
Lotfi Boubekeur and David.~H. Lyth.
\newblock {Hilltop inflation}.
\newblock {\em JCAP}, 07:010, 2005.

\bibitem{Martin:2006rs}
Jerome Martin and Christophe Ringeval.
\newblock {Inflation after WMAP3: Confronting the Slow-Roll and Exact Power
  Spectra to CMB Data}.
\newblock {\em JCAP}, 08:009, 2006.

\bibitem{Moss:2007qd}
IanG. Moss and ChristopherM. Graham.
\newblock {Testing models of inflation with CMB non-gaussianity}.
\newblock {\em JCAP}, 11:004, 2007.

\bibitem{Bezrukov:2010jz}
F.~Bezrukov, A.~Magnin, M.~Shaposhnikov, and S.~Sibiryakov.
\newblock {Higgs inflation: consistency and generalisations}.
\newblock {\em JHEP}, 01:016, 2011.

\bibitem{Zhao:2011zb}
Wen Zhao and Qing-Guo Huang.
\newblock {Testing inflationary consistency relations by the potential CMB
  observations}.
\newblock {\em Class. Quant. Grav.}, 28:235003, 2011.

\bibitem{Martin:2013nzq}
J\'er\^ome Martin, Christophe Ringeval, Roberto Trotta, and Vincent Vennin.
\newblock {The Best Inflationary Models After Planck}.
\newblock {\em JCAP}, 03:039, 2014.

\bibitem{Martin:2014rqa}
Jerome Martin, Christophe Ringeval, and Vincent Vennin.
\newblock {How Well Can Future CMB Missions Constrain Cosmic Inflation?}
\newblock {\em JCAP}, 10:038, 2014.

\bibitem{Martin:2014lra}
Jerome Martin, Christophe Ringeval, Roberto Trotta, and Vincent Vennin.
\newblock {Compatibility of Planck and BICEP2 in the Light of Inflation}.
\newblock {\em Phys. Rev. D}, 90(6):063501, 2014.

\bibitem{Carrillo-Gonzalez:2014tia}
Mariana Carrillo-Gonz\'alez, Gabriel Germ\'an-Velarde, Alfredo Herrera-Aguilar,
  Juan~Carlos Hidalgo, and Roberto Sussman.
\newblock {Testing Hybrid Natural Inflation with BICEP2}.
\newblock {\em Phys. Lett. B}, 734:345--349, 2014.

\bibitem{Creminelli:2014oaa}
Paolo Creminelli, Diana L\'opez~Nacir, Marko Simonovi\'c, Gabriele Trevisan,
  and Matias Zaldarriaga.
\newblock {$\phi^2$ or Not $\phi^2$: Testing the Simplest Inflationary
  Potential}.
\newblock {\em Phys. Rev. Lett.}, 112(24):241303, 2014.

\bibitem{DiValentino:2016nni}
Eleonora Di~Valentino and Laura Mersini-Houghton.
\newblock {Testing Predictions of the Quantum Landscape Multiverse 1: The
  Starobinsky Inflationary Potential}.
\newblock {\em JCAP}, 03:002, 2017.

\bibitem{DiValentino:2016ziq}
Eleonora Di~Valentino and Laura Mersini-Houghton.
\newblock {Testing Predictions of the Quantum Landscape Multiverse 2: The
  Exponential Inflationary Potential}.
\newblock {\em JCAP}, 03:020, 2017.

\bibitem{Campista:2017ovq}
Marcela Campista, Micol Benetti, and Jailson Alcaniz.
\newblock {Testing non-minimally coupled inflation with CMB data: a Bayesian
  analysis}.
\newblock {\em JCAP}, 09:010, 2017.

\bibitem{Giare:2019snj}
William Giar\`e, Eleonora Di~Valentino, and Alessandro Melchiorri.
\newblock {Testing the inflationary slow-roll condition with tensor modes}.
\newblock {\em Phys. Rev. D}, 99(12):123522, 2019.

\bibitem{Forconi:2021que}
Matteo Forconi, William Giar\`e, Eleonora Di~Valentino, and Alessandro
  Melchiorri.
\newblock {Cosmological constraints on slow roll inflation: An update}.
\newblock {\em Phys. Rev. D}, 104(10):103528, 2021.

\bibitem{Dai:2019ejv}
Rui Dai and Yi~Zhu.
\newblock {Testing kinetically coupled inflation models with CMB distortions}.
\newblock {\em JCAP}, 05:017, 2020.

\bibitem{Baumann:2015xxa}
Daniel Baumann, Hayden Lee, and Guilherme~L. Pimentel.
\newblock {High-Scale Inflation and the Tensor Tilt}.
\newblock {\em JHEP}, 01:101, 2016.

\bibitem{Odintsov:2020ilr}
S.~D. Odintsov, V.~K. Oikonomou, and F.~P. Fronimos.
\newblock {Canonical scalar field inflation with string and $R^2$
  -corrections}.
\newblock {\em Annals Phys.}, 424:168359, 2021.

\bibitem{Giare:2020plo}
William Giar\`e, Fabrizio Renzi, and Alessandro Melchiorri.
\newblock {Higher-Curvature Corrections and Tensor Modes}.
\newblock {\em Phys. Rev. D}, 103(4):043515, 2021.

\bibitem{Oikonomou:2021kql}
V.~K. Oikonomou.
\newblock {A refined Einstein\textendash{}Gauss\textendash{}Bonnet inflationary
  theoretical framework}.
\newblock {\em Class. Quant. Grav.}, 38(19):195025, 2021.

\bibitem{Odintsov:2022cbm}
Sergei~D. Odintsov, Vasilis~K. Oikonomou, and Ratbay Myrzakulov.
\newblock {Spectrum of Primordial Gravitational Waves in Modified Gravities: A
  Short Overview}.
\newblock {\em Symmetry}, 14(4):729, 2022.

\bibitem{Namba:2015gja}
Ryo Namba, Marco Peloso, Maresuke Shiraishi, Lorenzo Sorbo, and Caner Unal.
\newblock {Scale-dependent gravitational waves from a rolling axion}.
\newblock {\em JCAP}, 01:041, 2016.

\bibitem{Peloso:2016gqs}
Marco Peloso, Lorenzo Sorbo, and Caner Unal.
\newblock {Rolling axions during inflation: perturbativity and signatures}.
\newblock {\em JCAP}, 09:001, 2016.

\bibitem{Pi:2019ihn}
Shi Pi, Misao Sasaki, and Ying-li Zhang.
\newblock {Primordial Tensor Perturbation in Double Inflationary Scenario with
  a Break}.
\newblock {\em JCAP}, 06:049, 2019.

\bibitem{Ozsoy:2020ccy}
Ogan \"Ozsoy.
\newblock {Synthetic Gravitational Waves from a Rolling Axion Monodromy}.
\newblock {\em JCAP}, 04:040, 2021.

\bibitem{Stewart:2007fu}
Andrew Stewart and Robert Brandenberger.
\newblock {Observational Constraints on Theories with a Blue Spectrum of Tensor
  Modes}.
\newblock {\em JCAP}, 08:012, 2008.

\bibitem{Mukohyama:2014gba}
Shinji Mukohyama, Ryo Namba, Marco Peloso, and Gary Shiu.
\newblock {Blue Tensor Spectrum from Particle Production during Inflation}.
\newblock {\em JCAP}, 08:036, 2014.

\bibitem{Giovannini:2015kfa}
Massimo Giovannini.
\newblock {The refractive index of relic gravitons}.
\newblock {\em Class. Quant. Grav.}, 33(12):125002, 2016.

\bibitem{Giovannini:2018dob}
Massimo Giovannini.
\newblock {Post-inflationary thermal histories and the refractive index of
  relic gravitons}.
\newblock {\em Phys. Rev. D}, 98(10):103509, 2018.

\bibitem{Giovannini:2018nkt}
Massimo Giovannini.
\newblock {Blue and violet graviton spectra from a dynamical refractive index}.
\newblock {\em Phys. Lett. B}, 789:502--507, 2019.

\bibitem{Giovannini:2018zbf}
Massimo Giovannini.
\newblock {The propagating speed of relic gravitational waves and their
  refractive index during inflation}.
\newblock {\em Eur. Phys. J. C}, 78(6):442, 2018.

\bibitem{Giare:2020vhn}
William Giar\`e and Alessandro Melchiorri.
\newblock {Probing the inflationary background of gravitational waves from
  large to small scales}.
\newblock {\em Phys. Lett. B}, 815:136137, 2021.

\bibitem{Giare:2020vss}
William Giar\`e and Fabrizio Renzi.
\newblock {Propagating speed of primordial gravitational waves}.
\newblock {\em Phys. Rev. D}, 102(8):083530, 2020.

\bibitem{Giare:2022wxq}
William Giar\`e, Matteo Forconi, Eleonora Di~Valentino, and Alessandro
  Melchiorri.
\newblock {Towards a reliable calculation of relic radiation from primordial
  gravitational waves}.
\newblock {\em Mon. Not. Roy. Astron. Soc.}, 520:2, 2023.

\bibitem{Baumgart:2021ptt}
Matthew Baumgart, Jonathan~J. Heckman, and Logan Thomas.
\newblock {CFTs blueshift tensor fluctuations universally}.
\newblock {\em JCAP}, 07(07):034, 2022.

\bibitem{Franciolini:2018ebs}
G.~Franciolini, G.~F. Giudice, D.~Racco, and A.~Riotto.
\newblock {Implications of the detection of primordial gravitational waves for
  the Standard Model}.
\newblock {\em JCAP}, 05:022, 2019.

\bibitem{DEramo:2019tit}
Francesco D'Eramo and Kai Schmitz.
\newblock {Imprint of a scalar era on the primordial spectrum of gravitational
  waves}.
\newblock {\em Phys. Rev. Research.}, 1:013010, 2019.

\bibitem{Caldwell:2018giq}
Robert~R. Caldwell, Tristan~L. Smith, and Devin G.~E. Walker.
\newblock {Using a Primordial Gravitational Wave Background to Illuminate New
  Physics}.
\newblock {\em Phys. Rev. D}, 100(4):043513, 2019.

\bibitem{Clarke:2020bil}
Thomas~J. Clarke, Edmund~J. Copeland, and Adam Moss.
\newblock {Constraints on primordial gravitational waves from the Cosmic
  Microwave Background}.
\newblock {\em JCAP}, 10:002, 2020.

\bibitem{Caprini_2018}
Chiara Caprini and Daniel~G Figueroa.
\newblock Cosmological backgrounds of gravitational waves.
\newblock {\em CQG}, 35(16):163001, jul 2018.

\bibitem{Allen:1997ad}
Bruce Allen and Joseph~D. Romano.
\newblock {Detecting a stochastic background of gravitational radiation: Signal
  processing strategies and sensitivities}.
\newblock {\em Phys. Rev. D}, 59:102001, 1999.

\bibitem{Smith:2006nka}
Tristan~L. Smith, Elena Pierpaoli, and Marc Kamionkowski.
\newblock {A new cosmic microwave background constraint to primordial
  gravitational waves}.
\newblock {\em Phys. Rev. Lett.}, 97:021301, 2006.

\bibitem{Boyle:2007zx}
Latham~A. Boyle and Alessandra Buonanno.
\newblock {Relating gravitational wave constraints from primordial
  nucleosynthesis, pulsar timing, laser interferometers, and the CMB:
  Implications for the early Universe}.
\newblock {\em Phys. Rev. D}, 78:043531, 2008.

\bibitem{Kuroyanagi:2014nba}
Sachiko Kuroyanagi, Tomo Takahashi, and Shuichiro Yokoyama.
\newblock {Blue-tilted Tensor Spectrum and Thermal History of the Universe}.
\newblock {\em JCAP}, 02:003, 2015.

\bibitem{Ben-Dayan:2019gll}
Ido Ben-Dayan, Brian Keating, David Leon, and Ira Wolfson.
\newblock {Constraints on scalar and tensor spectra from $N_{eff}$}.
\newblock {\em JCAP}, 06:007, 2019.

\bibitem{Aich:2019obd}
Moumita Aich, Yin-Zhe Ma, Wei-Ming Dai, and Jun-Qing Xia.
\newblock {How much primordial tensor mode is allowed?}
\newblock {\em Phys. Rev. D}, 101(6):063536, 2020.

\bibitem{Cabass:2015jwe}
Giovanni Cabass, Luca Pagano, Laura Salvati, Martina Gerbino, Elena Giusarma,
  and Alessandro Melchiorri.
\newblock {Updated Constraints and Forecasts on Primordial Tensor Modes}.
\newblock {\em Phys. Rev. D}, 93(6):063508, 2016.

\bibitem{Vagnozzi:2020gtf}
Sunny Vagnozzi.
\newblock {Implications of the NANOGrav results for inflation}.
\newblock {\em Mon. Not. Roy. Astron. Soc.}, 502(1):L11--L15, 2021.

\bibitem{Benetti:2021uea}
Micol Benetti, Leila~Lobato Graef, and Sunny Vagnozzi.
\newblock {Primordial gravitational waves from NANOGrav: A broken power-law
  approach}.
\newblock {\em Phys. Rev. D}, 105(4):043520, 2022.

\bibitem{Calcagni:2020tvw}
Gianluca Calcagni and Sachiko Kuroyanagi.
\newblock {Stochastic gravitational-wave background in quantum gravity}.
\newblock {\em JCAP}, 03:019, 2021.

\bibitem{Oikonomou:2022ijs}
V.~K. Oikonomou.
\newblock {Amplification of the Primordial Gravitational Waves Energy Spectrum
  by a Kinetic Scalar in $F(R)$ Gravity}.
\newblock {\em Astropart. Phys.}, 144:102777, 2023.

\bibitem{Barrow:1993ad}
John~D. Barrow, J.~P. Mimoso, and M.~R. de~Garcia~Maia.
\newblock {Amplification of gravitational waves in scalar - tensor theories of
  gravity}.
\newblock {\em Phys. Rev. D}, 48:3630, 1993.
\newblock [Erratum: Phys.Rev.D 51, 5967 (1995)].

\bibitem{Peng:2021zon}
Zhi-Zhang Peng, Chengjie Fu, Jing Liu, Zong-Kuan Guo, and Rong-Gen Cai.
\newblock {Gravitational waves from resonant amplification of curvature
  perturbations during inflation}.
\newblock {\em JCAP}, 10:050, 2021.

\bibitem{Ota:2022hvh}
Atsuhisa Ota, Misao Sasaki, and Yi~Wang.
\newblock {Scale-invariant enhancement of gravitational waves during
  inflation}.
\newblock 9 2022.

\bibitem{Odintsov:2022sdk}
S.~D. Odintsov and V.~K. Oikonomou.
\newblock {Amplification of Primordial Gravitational Waves by a Geometrically
  Driven non-canonical Reheating Era}.
\newblock {\em Fortsch. Phys.}, 70(5):2100167, 2022.

\bibitem{Capurri:2020qgz}
Giulia Capurri, Nicola Bartolo, Davide Maino, and Sabino Matarrese.
\newblock {Let Effective Field Theory of Inflation flow: stochastic generation
  of models with red/blue tensor tilt}.
\newblock {\em JCAP}, 11:037, 2020.

\bibitem{Canas-Herrera:2021sjs}
Guadalupe Ca\~nas Herrera and Fabrizio Renzi.
\newblock {Current and future constraints on single-field
  \ensuremath{\alpha}-attractor models}.
\newblock {\em Phys. Rev. D}, 104(10):103512, 2021.

\bibitem{Odintsov:2023aaw}
S.~D. Odintsov, V.~K. Oikonomou, and F.~P. Fronimos.
\newblock {Inflationary Dynamics and Swampland Criteria for Modified
  Gauss-Bonnet Gravity Compatible with GW170817}.
\newblock 3 2023.

\bibitem{Oikonomou:2023bah}
V.~K. Oikonomou.
\newblock {Effects of the axion through the Higgs portal on primordial
  gravitational waves during the electroweak breaking}.
\newblock {\em Phys. Rev. D}, 107(6):064071, 2023.

\bibitem{Fronimos:2023tim}
F.~P. Fronimos and S.~A. Venikoudis.
\newblock {Inflationary phenomenology of non-minimally coupled
  Einstein-Chern-Simons gravity}.
\newblock 2 2023.

\bibitem{Cai:2022lec}
Yong Cai.
\newblock {Generating enhanced parity-violating gravitational waves during
  inflation with violation of the null energy condition}.
\newblock {\em Phys. Rev. D}, 107(6):063512, 2023.

\bibitem{Oikonomou:2022irx}
V.~K. Oikonomou.
\newblock {Effects of a pre-inflationary de Sitter bounce on the primordial
  gravitational waves in f(R) gravity theories}.
\newblock {\em Nucl. Phys. B}, 984:115985, 2022.

\bibitem{Gangopadhyay:2022vgh}
Mayukh~R. Gangopadhyay, Hussain~Ahmed Khan, and Yogesh.
\newblock {A case study of small field inflationary dynamics in the
  Einstein\textendash{}Gauss\textendash{}Bonnet framework in the light of
  GW170817}.
\newblock {\em Phys. Dark Univ.}, 40:101177, 2023.

\bibitem{Odintsov:2022hxu}
S.~D. Odintsov and V.~K. Oikonomou.
\newblock {Chirality of gravitational waves in Chern-Simons f(R) gravity
  cosmology}.
\newblock {\em Phys. Rev. D}, 105(10):104054, 2022.

\bibitem{Odintsov:2020mkz}
S.~D. Odintsov, V.~K. Oikonomou, F.~P. Fronimos, and S.~A. Venikoudis.
\newblock {GW170817-compatible constant-roll
  Einstein\textendash{}Gauss\textendash{}Bonnet inflation and
  non-Gaussianities}.
\newblock {\em Phys. Dark Univ.}, 30:100718, 2020.

\bibitem{Galloni:2022mok}
Giacomo Galloni, Nicola Bartolo, Sabino Matarrese, Marina Migliaccio, Angelo
  Ricciardone, and Nicola Vittorio.
\newblock {Updated constraints on amplitude and tilt of the tensor primordial
  spectrum}.
\newblock {\em JCAP}, 04:062, 2023.

\bibitem{Braglia:2022phb}
Matteo Braglia, Andrei Linde, Renata Kallosh, and Fabio Finelli.
\newblock {Hybrid \ensuremath{\alpha}-attractors, primordial black holes and
  gravitational wave backgrounds}.
\newblock {\em JCAP}, 04:033, 2023.

\bibitem{Giare:2023kiv}
William Giar\`e, Mariaveronica De~Angelis, Carsten van~de Bruck, and Eleonora
  Di~Valentino.
\newblock {Tracking the Multifield Dynamics with Cosmological Data: A Monte
  Carlo approach}.
\newblock 6 2023.

\bibitem{Antoniadis:2023zhi}
J.~Antoniadis et~al.
\newblock {The second data release from the European Pulsar Timing Array: V.
  Implications for massive black holes, dark matter and the early Universe}.
\newblock 6 2023.

\bibitem{NANOGrav:2023gor}
Gabriella Agazie et~al.
\newblock {The NANOGrav 15 yr Data Set: Evidence for a Gravitational-wave
  Background}.
\newblock {\em Astrophys. J. Lett.}, 951(1):L8, 2023.

\bibitem{NANOGrav:2023hvm}
Adeela Afzal et~al.
\newblock {The NANOGrav 15 yr Data Set: Search for Signals from New Physics}.
\newblock {\em Astrophys. J. Lett.}, 951(1):L11, 2023.

\bibitem{Vagnozzi:2023lwo}
Sunny Vagnozzi.
\newblock {Inflationary interpretation of the stochastic gravitational wave
  background signal detected by pulsar timing array experiments}.
\newblock 6 2023.

\bibitem{Oikonomou:2023qfz}
V.~K. Oikonomou.
\newblock {Flat Energy Spectrum of Primordial Gravitational Waves vs Peaks and
  the NANOGrav 2023 Observation}.
\newblock 6 2023.

\bibitem{BICEP:2021xfz}
P.~A.~R. Ade et~al.
\newblock {Improved Constraints on Primordial Gravitational Waves using Planck,
  WMAP, and BICEP/Keck Observations through the 2018 Observing Season}.
\newblock {\em Phys. Rev. Lett.}, 127(15):151301, 2021.

\bibitem{Lin:2019zdn}
Weikang Lin and Mustapha Ishak.
\newblock {A Bayesian interpretation of inconsistency measures in cosmology}.
\newblock {\em JCAP}, 05:009, 2021.

\bibitem{Handley:2020hdp}
Will Handley and Pablo Lemos.
\newblock {Quantifying the global parameter tensions between ACT, SPT and
  Planck}.
\newblock {\em Phys. Rev. D}, 103(6):063529, 2021.

\bibitem{LaPosta:2022llv}
Adrien La~Posta, Umberto Natale, Erminia Calabrese, Xavier Garrido, and Thibaut
  Louis.
\newblock {Assessing the consistency between CMB temperature and polarization
  measurements with application to Planck, ACT, and SPT data}.
\newblock {\em Phys. Rev. D}, 107(2):023510, 2023.

\bibitem{DiValentino:2022rdg}
Eleonora Di~Valentino, William Giar\`e, Alessandro Melchiorri, and Joseph Silk.
\newblock {Quantifying the global \textquoteleft{}CMB tension\textquoteright{}
  between the Atacama Cosmology Telescope and the Planck satellite in extended
  models of cosmology}.
\newblock {\em Mon. Not. Roy. Astron. Soc.}, 520(1):210--215, 2023.

\bibitem{DiValentino:2022oon}
Eleonora Di~Valentino, William Giar\`e, Alessandro Melchiorri, and Joseph Silk.
\newblock {Health checkup test of the standard cosmological model in view of
  recent cosmic microwave background anisotropies experiments}.
\newblock {\em Phys. Rev. D}, 106(10):103506, 2022.

\bibitem{Giare:2022rvg}
William Giar\`e, Fabrizio Renzi, Olga Mena, Eleonora Di~Valentino, and
  Alessandro Melchiorri.
\newblock {Is the Harrison-Zel'dovich spectrum coming back? ACT preference for
  $n_s \sim 1$ and its discordance with Planck}.
\newblock 10 2022.

\bibitem{Calderon:2023obf}
Rodrigo Calder\'on, Arman Shafieloo, Dhiraj~Kumar Hazra, and Wuhyun Sohn.
\newblock {On the consistency of $\Lambda$CDM with CMB measurements in light of
  the latest Planck, ACT, and SPT data}.
\newblock 2 2023.

\bibitem{Giare:2023xoc}
William Giar\`e.
\newblock {CMB Anomalies and the Hubble Tension}.
\newblock 5 2023.

\bibitem{Amoros:2014tha}
Jaume Amor\'os, Jaume de~Haro, and Sergei~D. Odintsov.
\newblock {$R+\alpha R^2$ Loop Quantum Cosmology}.
\newblock {\em Phys. Rev. D}, 89(10):104010, 2014.

\bibitem{Kallosh:2013yoa}
Renata Kallosh, Andrei Linde, and Diederik Roest.
\newblock {Superconformal Inflationary $\alpha$-Attractors}.
\newblock {\em JHEP}, 11:198, 2013.

\bibitem{Kehagias:2013mya}
Alex Kehagias, Azadeh Moradinezhad~Dizgah, and Antonio Riotto.
\newblock {Remarks on the Starobinsky model of inflation and its descendants}.
\newblock {\em Phys. Rev. D}, 89(4):043527, 2014.

\bibitem{Peebles:1998qn}
P.~J.~E. Peebles and A.~Vilenkin.
\newblock {Quintessential inflation}.
\newblock {\em Phys. Rev. D}, 59:063505, 1999.

\bibitem{Dimopoulos:2017zvq}
Konstantinos Dimopoulos and Charlotte Owen.
\newblock {Quintessential Inflation with $\alpha$-attractors}.
\newblock {\em JCAP}, 06:027, 2017.

\bibitem{AresteSalo:2021wgb}
Llibert Arest\'e~Sal\'o, David Benisty, Eduardo~I. Guendelman, and Jaume
  de~Haro.
\newblock {$\alpha$-attractors in quintessential inflation motivated by
  supergravity}.
\newblock {\em Phys. Rev. D}, 103(12):123535, 2021.

\bibitem{deHaro:2016cdm}
Jaume de~Haro, Jaume Amor\'{o}s, and Supriya Pan.
\newblock {Simple inflationary quintessential model II: Power law potentials}.
\newblock {\em Phys. Rev. D}, 94(6):064060, 2016.

\bibitem{Lewis:1999bs}
Antony Lewis, Anthony Challinor, and Anthony Lasenby.
\newblock {Efficient computation of CMB anisotropies in closed FRW models}.
\newblock {\em Astrophys. J.}, 538:473--476, 2000.

\bibitem{Howlett:2012mh}
Cullan Howlett, Antony Lewis, Alex Hall, and Anthony Challinor.
\newblock {CMB power spectrum parameter degeneracies in the era of precision
  cosmology}.
\newblock {\em JCAP}, 04:027, 2012.

\bibitem{Torrado:2020xyz}
Jesus Torrado and Antony Lewis.
\newblock {Cobaya: Code for Bayesian Analysis of hierarchical physical models}.
\newblock {\em arXiv:2005.05290}, 5 2020.

\bibitem{Lewis:2002ah}
Antony Lewis and Sarah Bridle.
\newblock {Cosmological parameters from CMB and other data: A Monte Carlo
  approach}.
\newblock {\em Phys. Rev. D}, 66:103511, 2002.

\bibitem{Neal:2005}
R.~M. {Neal}.
\newblock {Taking Bigger Metropolis Steps by Dragging Fast Variables}.
\newblock {\em ArXiv Mathematics e-prints}, February 2005.

\bibitem{Gelman:1992zz}
Andrew Gelman and Donald~B. Rubin.
\newblock {Inference from Iterative Simulation Using Multiple Sequences}.
\newblock {\em Statist. Sci.}, 7:457--472, 1992.

\bibitem{Planck:2018vyg}
N.~Aghanim et~al.
\newblock {Planck 2018 results. VI. Cosmological parameters}.
\newblock {\em Astron. Astrophys.}, 641:A6, 2020.
\newblock [Erratum: Astron.Astrophys. 652, C4 (2021)].

\bibitem{Planck:2018lbu}
N.~Aghanim et~al.
\newblock {Planck 2018 results. VIII. Gravitational lensing}.
\newblock {\em Astron. Astrophys.}, 641:A8, 2020.

\bibitem{BOSS:2012dmf}
Kyle~S. Dawson et~al.
\newblock {The Baryon Oscillation Spectroscopic Survey of SDSS-III}.
\newblock {\em Astron. J.}, 145:10, 2013.

\bibitem{Dawson:2015wdb}
Kyle~S. Dawson et~al.
\newblock {The SDSS-IV extended Baryon Oscillation Spectroscopic Survey:
  Overview and Early Data}.
\newblock {\em Astron. J.}, 151:44, 2016.

\bibitem{Heavens:2017hkr}
Alan Heavens, Yabebal Fantaye, Elena Sellentin, Hans Eggers, Zafiirah Hosenie,
  Steve Kroon, and Arrykrishna Mootoovaloo.
\newblock {No evidence for extensions to the standard cosmological model}.
\newblock {\em Phys. Rev. Lett.}, 119(10):101301, 2017.

\bibitem{Heavens:2017afc}
Alan Heavens, Yabebal Fantaye, Arrykrishna Mootoovaloo, Hans Eggers, Zafiirah
  Hosenie, Steve Kroon, and Elena Sellentin.
\newblock {Marginal Likelihoods from Monte Carlo Markov Chains}.
\newblock {\em arXiv:1704.03472 [stat.CO]}, 4 2017.

\bibitem{Trotta:2008qt}
Roberto Trotta.
\newblock {Bayes in the sky: Bayesian inference and model selection in
  cosmology}.
\newblock {\em Contemp. Phys.}, 49:71--104, 2008.

\bibitem{BICEP3}
J.~A. Grayson et~al.
\newblock {BICEP3 performance overview and planned Keck Array upgrade}.
\newblock {\em Proc. SPIE Int. Soc. Opt. Eng.}, 9914:99140S, 2016.

\bibitem{CLASS}
Thomas Essinger-Hileman et~al.
\newblock {CLASS: The Cosmology Large Angular Scale Surveyor}.
\newblock {\em Proc. SPIE Int. Soc. Opt. Eng.}, 9153:91531I, 2014.

\bibitem{LBIRD}
A.~Suzuki et~al.
\newblock The litebird satellite mission: Sub-kelvin instrument.
\newblock {\em Journal of Low Temperature Physics}, 193(5):1048--1056, Dec
  2018.

\bibitem{Abazajian:2019eic}
Kevork Abazajian et~al.
\newblock {CMB-S4 Science Case, Reference Design, and Project Plan}.
\newblock 7 2019.

\bibitem{CMB-S4}
Kevork~N. Abazajian et~al.
\newblock {CMB-S4 Science Book, First Edition}.
\newblock 2016.

\bibitem{SimonsObservatory:2018koc}
Peter Ade et~al.
\newblock {The Simons Observatory: Science goals and forecasts}.
\newblock {\em JCAP}, 02:056, 2019.

\bibitem{NASAPICO:2019thw}
Shaul Hanany et~al.
\newblock {PICO: Probe of Inflation and Cosmic Origins}.
\newblock 3 2019.

\bibitem{CMB-HD:2022bsz}
Simone Aiola et~al.
\newblock {Snowmass2021 CMB-HD White Paper}.
\newblock 3 2022.

\bibitem{Lidsey:1995np}
James~E. Lidsey, Andrew~R. Liddle, Edward~W. Kolb, Edmund~J. Copeland, Tiago
  Barreiro, and Mark Abney.
\newblock {Reconstructing the inflation potential : An overview}.
\newblock {\em Rev. Mod. Phys.}, 69:373--410, 1997.

\bibitem{Liddle:2003as}
Andrew~R Liddle and Samuel~M Leach.
\newblock {How long before the end of inflation were observable perturbations
  produced?}
\newblock {\em Phys. Rev. D}, 68:103503, 2003.

\bibitem{deHaro:2023xcc}
Jaume de~Haro.
\newblock {Reheating constraints in instant preheating}.
\newblock {\em Phys. Rev. D}, 107(12):123511, 2023.

\bibitem{DiValentino:2018zjj}
Eleonora Di~Valentino, Alessandro Melchiorri, Yabebal Fantaye, and Alan
  Heavens.
\newblock {Bayesian evidence against the Harrison-Zel\textquoteright{}dovich
  spectrum in tensions with cosmological data sets}.
\newblock {\em Phys. Rev. D}, 98(6):063508, 2018.

\bibitem{Ye:2022efx}
Gen Ye, Jun-Qian Jiang, and Yun-Song Piao.
\newblock {Toward inflation with ns=1 in light of the Hubble tension and
  implications for primordial gravitational waves}.
\newblock {\em Phys. Rev. D}, 106(10):103528, 2022.

\bibitem{Jiang:2022uyg}
Jun-Qian Jiang and Yun-Song Piao.
\newblock {Toward early dark energy and $n_s=1$ with Planck, ACT, and SPT
  observations}.
\newblock {\em Phys. Rev. D}, 105(10):103514, 2022.

\bibitem{Jiang:2022qlj}
Jun-Qian Jiang, Gen Ye, and Yun-Song Piao.
\newblock {Return of Harrison-Zeldovich spectrum in light of recent
  cosmological tensions}.
\newblock 10 2022.

\bibitem{Takahashi:2021bti}
Fuminobu Takahashi and Wen Yin.
\newblock {Cosmological implications of ns~1 in light of the Hubble tension}.
\newblock {\em Phys. Lett. B}, 830:137143, 2022.

\bibitem{Lin:2022gbl}
Chia-Min Lin.
\newblock {D-term inflation in braneworld models: Consistency with
  cosmic-string bounds and early-time Hubble tension resolving models}.
\newblock {\em Phys. Rev. D}, 106(10):103511, 2022.

\bibitem{Hazra:2022rdl}
Dhiraj~Kumar Hazra, Akhil Antony, and Arman Shafieloo.
\newblock {One spectrum to cure them all: signature from early Universe solves
  major anomalies and tensions in cosmology}.
\newblock {\em JCAP}, 08(08):063, 2022.

\bibitem{Braglia:2021sun}
Matteo Braglia, Xingang Chen, and Dhiraj~Kumar Hazra.
\newblock {Uncovering the history of cosmic inflation from anomalies in cosmic
  microwave background spectra}.
\newblock {\em Eur. Phys. J. C}, 82(5):498, 2022.

\bibitem{Keeley:2020rmo}
Ryan~E. Keeley, Arman Shafieloo, Dhiraj~Kumar Hazra, and Tarun Souradeep.
\newblock {Inflation Wars: A New Hope}.
\newblock {\em JCAP}, 09:055, 2020.

\bibitem{Jiang:2023bsz}
Jun-Qian Jiang, Gen Ye, and Yun-Song Piao.
\newblock {Impact of the Hubble tension on the $r$-$n_s$ contour}.
\newblock 3 2023.

\bibitem{Peng:2023bik}
Ze-Yu Peng and Yun-Song Piao.
\newblock {Testing the $n_s-H_0$ scaling relation with Planck-independent CMB
  data}.
\newblock 8 2023.

\bibitem{Riess:2021jrx}
Adam~G. Riess et~al.
\newblock {A Comprehensive Measurement of the Local Value of the Hubble
  Constant with 1 km/s/Mpc Uncertainty from the Hubble Space Telescope and the
  SH0ES Team}.
\newblock {\em Astrophys. J. Lett.}, 934(1):L7, 2022.

\bibitem{DiValentino:2021izs}
Eleonora Di~Valentino, Olga Mena, Supriya Pan, Luca Visinelli, Weiqiang Yang,
  Alessandro Melchiorri, David~F. Mota, Adam~G. Riess, and Joseph Silk.
\newblock {In the realm of the Hubble tension\textemdash{}a review of
  solutions}.
\newblock {\em Class. Quant. Grav.}, 38(15):153001, 2021.

\bibitem{Perivolaropoulos:2021jda}
Leandros Perivolaropoulos and Foteini Skara.
\newblock {Challenges for \ensuremath{\Lambda}CDM: An update}.
\newblock {\em New Astron. Rev.}, 95:101659, 2022.

\bibitem{Schoneberg:2021qvd}
Nils Sch\"oneberg, Guillermo Franco~Abell\'an, Andrea P\'erez~S\'anchez,
  Samuel~J. Witte, Vivian Poulin, and Julien Lesgourgues.
\newblock {The $H_0$ Olympics: A fair ranking of proposed models}.
\newblock {\em Phys. Rept.}, 984:1--55, 2022.

\bibitem{Abdalla:2022yfr}
Elcio Abdalla et~al.
\newblock {Cosmology intertwined: A review of the particle physics,
  astrophysics, and cosmology associated with the cosmological tensions and
  anomalies}.
\newblock {\em JHEAp}, 34:49--211, 2022.

\bibitem{Kallosh:2022ggf}
Renata Kallosh and Andrei Linde.
\newblock {Hybrid cosmological attractors}.
\newblock {\em Phys. Rev. D}, 106(2):023522, 2022.

\bibitem{Calabrese:2011hg}
Erminia Calabrese, Dragan Huterer, Eric~V. Linder, Alessandro Melchiorri, and
  Luca Pagano.
\newblock {Limits on Dark Radiation, Early Dark Energy, and Relativistic
  Degrees of Freedom}.
\newblock {\em Phys. Rev. D}, 83:123504, 2011.

\bibitem{Poulin:2018dzj}
Vivian Poulin, Tristan~L. Smith, Daniel Grin, Tanvi Karwal, and Marc
  Kamionkowski.
\newblock {Cosmological implications of ultralight axionlike fields}.
\newblock {\em Phys. Rev. D}, 98(8):083525, 2018.

\bibitem{Niedermann:2019olb}
Florian Niedermann and Martin~S. Sloth.
\newblock {New Early Dark Energy}.
\newblock 10 2019.

\bibitem{Niedermann:2020dwg}
Florian Niedermann and Martin~S. Sloth.
\newblock {Resolving the Hubble Tension with New Early Dark Energy}.
\newblock 6 2020.

\bibitem{Murgia:2020ryi}
Riccardo Murgia, Guillermo~F. Abell\'an, and Vivian Poulin.
\newblock {Early dark energy resolution to the Hubble tension in light of weak
  lensing surveys and lensing anomalies}.
\newblock {\em Phys. Rev. D}, 103(6):063502, 2021.

\bibitem{Herold:2021ksg}
Laura Herold, Elisa G.~M. Ferreira, and Eiichiro Komatsu.
\newblock {New Constraint on Early Dark Energy from Planck and BOSS Data Using
  the Profile Likelihood}.
\newblock {\em Astrophys. J. Lett.}, 929(1):L16, 2022.

\bibitem{Reeves:2022aoi}
Alexander Reeves, Laura Herold, Sunny Vagnozzi, Blake~D. Sherwin, and Elisa
  G.~M. Ferreira.
\newblock {Restoring cosmological concordance with early dark energy and
  massive neutrinos?}
\newblock 7 2022.

\bibitem{Niedermann:2023ssr}
Florian Niedermann and Martin~S. Sloth.
\newblock {New Early Dark Energy as a solution to the $H_0$ and $S_8$
  tensions}.
\newblock 7 2023.

\bibitem{Cruz:2023lmn}
Juan~S. Cruz, Florian Niedermann, and Martin~S. Sloth.
\newblock {Cold New Early Dark Energy pulls the trigger on the $H_0$ and $S_8$
  tensions: a simultaneous solution to both tensions without new ingredients}.
\newblock 5 2023.

\bibitem{Eskilt:2023nxm}
Johannes~R. Eskilt, Laura Herold, Eiichiro Komatsu, Kai Murai, Toshiya
  Namikawa, and Fumihiro Naokawa.
\newblock {Constraint on Early Dark Energy from Isotropic Cosmic
  Birefringence}.
\newblock 3 2023.

\bibitem{Poulin:2018cxd}
Vivian Poulin, Tristan~L. Smith, Tanvi Karwal, and Marc Kamionkowski.
\newblock {Early Dark Energy Can Resolve The Hubble Tension}.
\newblock {\em Phys. Rev. Lett.}, 122(22):221301, 2019.

\bibitem{Hill:2021yec}
J.~Colin Hill et~al.
\newblock {Atacama Cosmology Telescope: Constraints on prerecombination early
  dark energy}.
\newblock {\em Phys. Rev. D}, 105(12):123536, 2022.

\bibitem{Poulin:2023lkg}
Vivian Poulin, Tristan~L. Smith, and Tanvi Karwal.
\newblock {The Ups and Downs of Early Dark Energy solutions to the Hubble
  tension: a review of models, hints and constraints circa 2023}.
\newblock 2 2023.

\bibitem{Bassett:2005xm}
Bruce~A. Bassett, Shinji Tsujikawa, and David Wands.
\newblock {Inflation dynamics and reheating}.
\newblock {\em Rev. Mod. Phys.}, 78:537--589, 2006.

\end{thebibliography}



\end{document}